\documentclass[aps,pra,onecolumn,superscriptaddress,groupedaddress,nofootinbib,amsart]{revtex4}

\usepackage[dvipsnames]{xcolor}
\usepackage{natbib}
\usepackage{amssymb,amsmath,latexsym}
\usepackage{amsthm}
\usepackage{hyperref}
\usepackage{bm}
\usepackage{microtype}
\usepackage{url}
\usepackage{graphicx}
\usepackage{listings}
\usepackage{xspace}
\usepackage{cleveref}
\usepackage[T1]{fontenc}
\usepackage{pgf-umlsd}
\usepackage{caption}
\usepackage{subcaption}
\usepackage[section]{placeins}

\newcommand{\dropcap}[1]{#1}

\theoremstyle{plain}

\theoremstyle{definition}

\theoremstyle{remark}

\usepackage{xargs}
\usepackage[colorinlistoftodos,prependcaption]{todonotes}
\newcommandx{\unsure}[2][1=]{\todo[linecolor=red,backgroundcolor=red!25,bordercolor=red,#1]{#2}}
\newcommandx{\change}[2][1=]{\todo[linecolor=blue,backgroundcolor=blue!25,bordercolor=blue,#1]{#2}}
\newcommandx{\note}[2][1=]{\todo[linecolor=green,backgroundcolor=green!25,bordercolor=green,#1]{#2}}
\newcommandx{\improvement}[2][1=]{\todo[linecolor=purple,backgroundcolor=purple!25,bordercolor=purple,#1]{#2}}
\newcommandx{\thiswillnotshow}[2][1=]{\todo[disable,#1]{#2}}

\graphicspath{{./figs/}}

\usepackage{tikz}
\usetikzlibrary{shapes.geometric, arrows, fit, calc,  positioning, backgrounds, fadings, arrows.meta}

\usepackage{enumitem}

\definecolor{codegreen}{rgb}{0,0.6,0}
\definecolor{codegray}{rgb}{0.5,0.5,0.5}
\definecolor{codepurple}{rgb}{0.58,0,0.82}
\definecolor{backcolour}{rgb}{0.95,0.95,0.92}

\lstdefinestyle{mystyle}{
    commentstyle=\color{codegreen},
    keywordstyle=\color{magenta},
    numberstyle=\tiny\color{codegray},
    stringstyle=\color{codepurple},
    basicstyle=\ttfamily\footnotesize,
    breakatwhitespace=false,
    breaklines=true,
    captionpos=b,
    keepspaces=true,
    numbers=left,
    firstnumber=0,
    numbersep=5pt,
    showspaces=false,
    showstringspaces=false,
    showtabs=false,
    tabsize=2
}
\lstdefinestyle{inlinestyle}{
    commentstyle=\color{codegreen},
    keywordstyle=\color{magenta},
    numberstyle=\tiny\color{codegray},
    stringstyle=\color{codepurple},
    basicstyle=\ttfamily\footnotesize\color{PineGreen},
    breakatwhitespace=false,
    breaklines=true,
    captionpos=b,
    keepspaces=true,
    numbers=left,
    firstnumber=0,
    numbersep=5pt,
    showspaces=false,
    showstringspaces=false,
    showtabs=false,
    tabsize=2
}
\lstdefinelanguage{netqasm}
{
    morekeywords={
        add, add, sub, addm, subm,
        jmp, bez, bnz, beq, bne, blt, bge,
        set, store, load, undef, lea,
        array, qalloc, qfree,
        wait_all, wait_any, wait_single,
        ret_reg, ret_arr,
        meas,
        create_epr, recv_epr,
        init, x, y, z, h, s, k, t, rot_x, rot_y, rot_z, cnot, cphase, cx_dir, cy_dir,
        APPID, NETQASM,
    }
    sensitive=false,
    morecomment=[l]{//},
    morecomment=[s]{/*}{*/},
    morecomment=[s][\color{blue}]{\#\ }{\ },
    morestring=[b]",
}

\lstnewenvironment{nqcode}{\lstset{style=mystyle,language=netqasm}}{}
\lstnewenvironment{pycode}{\lstset{style=mystyle,language=Python}}{}
\newcommand{\nq}[1]{\lstinline[style=inlinestyle,language=C++]{#1}}
\newcommand{\py}[1]{\lstinline[language=Python]{#1}}
\usepackage[most]{tcolorbox}
\usepackage{etoolbox}
\tcbset{
}
\BeforeBeginEnvironment{nqcode}{\begin{tcolorbox}[colback=white]}%
\AfterEndEnvironment{nqcode}{\end{tcolorbox}}%
\BeforeBeginEnvironment{pycode}{\begin{tcolorbox}[colback=white]}%
\AfterEndEnvironment{pycode}{\end{tcolorbox}}%

\newcommand{\netqasm}{\nq{NetQASM}\xspace}
\newcommand{\qasm}{\nq{QASM}\xspace}
\newcommand{\openqasm}{\nq{OpenQASM}\xspace}
\newcommand{\cqc}{\nq{CQC}\xspace}
\newcommand{\qnodeos}{\nq{QNodeOS}\xspace}

\newcommand{\host}{application layer\xspace}
\newcommand{\QNPU}{\nq{QNPU}\xspace}
\newcommand{\netsquid}{\nq{NetSquid}\xspace}
\newcommand{\squidasm}{\nq{SquidASM}\xspace}
\newcommand{\simulaqron}{\nq{SimulaQron}\xspace}
\newcommand{\IMMEDIATE}{\textbf{IMMEDIATE}\xspace}
\newcommand{\REGISTER}{\textbf{REGISTER}\xspace}
\newcommand{\ADDRESS}{\textbf{ADDRESS}\xspace}

\begin{document}

\title{NetQASM - A low-level instruction set architecture for hybrid quantum-classical programs in a quantum internet}

\author{Axel Dahlberg}
\thanks{These authors contributed equally.}
\affiliation{QuTech, Lorentzweg 1, 2628 CJ Delft, Netherlands}
\email[]{Contact: b.vandervecht@tudelft.nl, s.d.c.wehner@tudelft.nl}
\affiliation{Kavli Institute of Nanoscience, Delft, The Netherlands}

\author{Bart van der Vecht}
\thanks{These authors contributed equally.}
\affiliation{QuTech, Lorentzweg 1, 2628 CJ Delft, Netherlands}
\affiliation{Kavli Institute of Nanoscience, Delft, The Netherlands}

\author{Carlo Delle Donne}
\affiliation{QuTech, Lorentzweg 1, 2628 CJ Delft, Netherlands}
\affiliation{Kavli Institute of Nanoscience, Delft, The Netherlands}

\author{Matthew Skrzypczyk}
\affiliation{QuTech, Lorentzweg 1, 2628 CJ Delft, Netherlands}
\affiliation{Kavli Institute of Nanoscience, Delft, The Netherlands}

\author{Ingmar te Raa}
\affiliation{QuTech, Lorentzweg 1, 2628 CJ Delft, Netherlands}
\affiliation{Kavli Institute of Nanoscience, Delft, The Netherlands}

\author{Wojciech Kozlowski}
\affiliation{QuTech, Lorentzweg 1, 2628 CJ Delft, Netherlands}
\affiliation{Kavli Institute of Nanoscience, Delft, The Netherlands}

\author{Stephanie Wehner}
\affiliation{QuTech, Lorentzweg 1, 2628 CJ Delft, Netherlands}
\affiliation{Kavli Institute of Nanoscience, Delft, The Netherlands}
\email[]{s.d.c.wehner@tudelft.nl}

\begin{abstract}
    We introduce NetQASM, a low-level instruction set architecture for quantum
    internet applications. NetQASM is a universal, platform-independent and
    extendable instruction set with support for local quantum gates, powerful
    classical logic and quantum networking operations for remote entanglement
    generation. Furthermore, NetQASM allows for close integration of classical
    logic and communication at the application layer with quantum operations at
    the physical layer. This enables quantum network applications to be
    programmed in high-level platform-independent software, which is not
    possible using any other QASM variants. We implement NetQASM in a series of
    tools to write, parse, encode and run NetQASM code, which are available
    online. Our tools include a higher-level SDK in Python, which allows an easy
    way of programming applications for a quantum internet. Our SDK can be used
    at home by making use of our existing quantum simulators, NetSquid and
    SimulaQron, and will also provide a public interface to hardware released on
    a future iteration of Quantum Network Explorer.
\end{abstract}

\maketitle

\section{Introduction}\label{sec:introduction} \dropcap{Q}uantum mechanics shows
that if one is able to communicate quantum information between nodes in a
network, one is able to achieve certain tasks which are impossible using only
classical communication. There are many applications~\cite{Wehner2018stages}
where a \emph{quantum network} has advantage over a \emph{classical
      (non-quantum) network}, either by (1) enabling something that is theoretically
impossible in a classical network, such as the establishment of an
unconditionally secure key~\cite{bb84} and secure blind quantum
computing~\cite{childs2005assisted} or (2) allowing something to be done faster
or more efficiently such as exponential savings in
communication~\cite{Buhrman2010} and extending the baseline of
telescopes~\cite{gottesman2012longer}. In recent years, many experiments have
been conducted to show that a quantum network is not only a theoretical concept,
and indeed advancements have been made to implement such a quantum network on
various hardware platforms. \cite{Hensen2015, Humphreys2018,
      moehring2007entanglement, hofmann2012heralded, Kalb2017, Inlek2017,
      sangouard2011quantum}. However, these experiments alone do not yet make a
quantum network \textit{programmable}, since the program logic was hard-coded
into the experimental hardware ahead of time.\footnote{There have been examples of
      experiments with some simple logic but only with a very limited number of
      pre-loaded decision-branches.}

Before considering how to program quantum network applications, let us first
briefly sketch the system our applications are run on. Abstractly, quantum
networks consist of \textit{nodes} that are connected by \textit{channels}
(\cref{fig:network_model}). Classical channels enable classical communication
between nodes, while quantum channels are used for \textit{entanglement}
generation between nodes. So-called \textit{end-nodes} may contain
\textit{quantum processors} that can run arbitrary (quantum) programs. They have
access to a quantum memory consisting of qubits, on which they can perform
operations, including quantum computations. Some of these qubits may be used for
establishing an entangled quantum state with a remote node. An end-node also
possesses a classical processor and a classical memory. Furthermore, an end-node
can send and receive classical messages to and from other end-nodes in the
network. A network of quantum networks may be a called a \textit{quantum
      internet}.

Quantum (network) processors differ from classical processors in a number of
ways. Firstly, quantum memory has limited lifetime, meaning that its quality
degrades over time. For example, quantum memories based on nitrogen-vacancy (NV)
centers in diamond have impressively been optimized to achieve lifetimes in the
order of seconds~\cite{Abobeih2018}; however, this is still very short compared
to classical memories, which generally do not have a limited lifetime at all.
Therefore, the quality of program execution is time-sensitive. Secondly,
physical devices are prone to inaccuracies which lead to decreased quality of
(quantum) computation. For example, applying an operation (like a gate) on a
qubit affects that qubit's quality. We note that the two challenges mentioned so
far are also inherent to non-network quantum processors. Quantum
\textit{network} processors have additional challenges: (1) the processor may
have to act as a local computation unit and a network interface at the same
time; for example, in NV centers, an electron spin qubit is used for generating
entanglement with a remote node but is also needed to do local two-qubit gates,
(2) remote-entanglement operations may not have a fixed time in which they
complete, which makes scheduling and optimization more difficult.

\begin{figure}[h]
      \centering
      \includegraphics[width=0.6\textwidth]{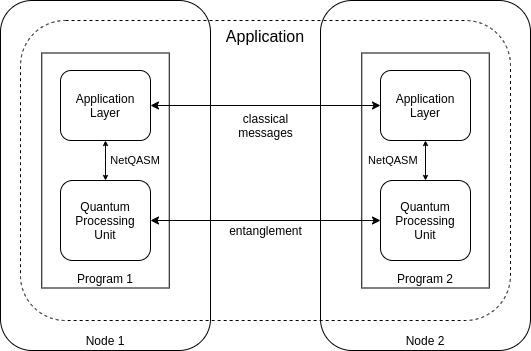}
      \caption{A quantum network application consists of a program for each of
            the nodes involved in the application. Each program is locally
            executed by the node. Program execution on each node is split into
            execution in an application layer, which can send and receive
            classical messages, and a quantum processor, which can create
            entanglement with another node. The communication between nodes can
            hence be both classical and quantum. Communication instructions need
            to be matched by corresponding instructions in the other program.
            There is no global actor overseeing execution of each of the
            programs, and the nodes may be physically far
            apart.}\label{fig:app_programs}
\end{figure}

Quantum network \textit{applications}, also called \textit{protocols}, are
multi-partite programs that involve entanglement generation and classical
communication between different end-nodes, as well as local computation.
Examples include Quantum Key Distribution (QKD)~\cite{bb84, ekert1991quantum},
leader election protocols~\cite{kobayashi2014simpler, ganz2009quantum}, and
Blind Quantum Compuation (BQC)~\cite{Wehner2018stages}. Such applications are
split into distinct \textit{programs} each of which runs on a separate end-node.
The programs consist of both local operations (classical and quantum) and
network operations (classical and quantum), see \cref{fig:app_programs}. That
is, the programs communicate either by passing classical messages, or by
establishing quantum entanglement. For example, BQC involves a \textit{client}
node and a \textit{server} node, both of which run their own program. Their
joint execution looks roughly as follows: (1) The client and server engage in
remote entanglement generation such that the server's quantum memory ends up
being in a certain state, (2) the client sends instructions to the server in the
form of a classical message, (3) the server performs a measurement-based
computation on its own quantum memory based on the client's instructions, (4)
the server sends measurement results back to the client, (5) the client sends
new instructions based on the measurement results, (6) repeat steps 3 to 5 until
the client obtains its desired result.

\begin{figure}[h]
      \centering
      \includegraphics[width=0.6\textwidth]{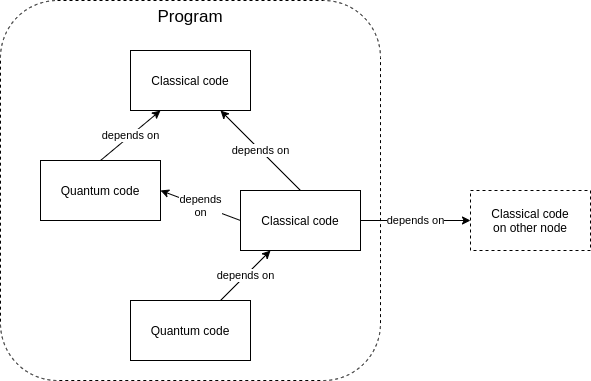}
      \caption{A program on a single node consists of different blocks of code,
            which can be quantum (pure quantum instructions with classical
            control in between), or classical (no quantum operations at all).
            These blocks may depend on each other in various ways. For example,
            the outcome of a measurement happening in one of the quantum blocks
            may be used in a calculation performed in one of the classical
            blocks. Blocks may also depend on other nodes. For instance, the
            value of a message coming from another node can influence the branch
            taken in one of the classical blocks.}\label{fig:program_decomp}
\end{figure}

The example above illustrates that quantum network programs consist of different
types of operations. Indeed, program code consists of \textit{classical code},
containing local classical operations and classical communication with other
nodes, and \textit{quantum code}, which are operations on quantum memory (such
as \textit{gates}) and remote entanglement generation. Blocks of these types of
code may depend on each other in multiple ways, as depicted
in~\cref{fig:program_decomp}. Programs with mixed classical and quantum
operations have also been called \textit{dynamic quantum
      circuits}~\cite{cross2021openqasm, burgholzer2021towards}, but these do not
cover the networking dimension found in programs we consider here, such as the
dependency on remote information and entanglement generation operations.

Due to the nature of quantum network programs, execution may have to
\textit{wait} for some time. For example, the program needs to wait until
another node sends a classical message, or until remote entanglement has been
established. Therefore, it makes sense to run multiple (independent) quantum
network programs on a node at the same time (interleaved), so that processor
idle times can be filled by execution of other programs. This is something that
typically does not happen on local quantum computers, and therefore introduces
new challenges.

Quantum network applications may be programmed by a single actor. For example, a
developer may program a QKD application in the form of a two programs, and
distribute these two programs to two end-nodes in the network. Alternatively, a
single-node quantum network program may be developed separately from other
programs, possibly not knowing how these other programs are implemented. For
example, a BQC service provider could have already implemented the server-side
program of a specific BQC protocol. A client may then write the client-side of
this protocol, without having control over the server-side implementation.

The aim of this work is to propose a way to program quantum network programs
and execute them on the end-nodes of a quantum network.

\subsection{Contribution}
In this work we introduce an abstract model---including a quantum network processing unit (\QNPU)--- for end-nodes in a quantum network,
which we define in~\cref{sec:preliminaries}. We then propose \netqasm, an
instruction set architecture that can be used to run arbitrary programs (of the form described in~\cref{fig:program_decomp}) on
end-nodes, as long as the end-nodes realize the model including the QNPU.

\netqasm consists of a specification of a low-level assembly-like
language to express the quantum parts of quantum network program code.
It also specifies how the application layer should interact with the \QNPU and
how the assembly language can be used to execute (network) quantum code.
This is not possible using other \qasm languages.

The \netqasm language is extendible using the concept of \textit{flavors}. The
core language definition consists of a common set of instructions that are
shared by all flavors. This common set contains classical instructions for
control-flow and classical memory operations. This allows the realization of
low-level control logic close to the quantum hardware; for example, to perform
branching based on a measurement outcome. Quantum-specific instructions are
bundled in flavors. We introduce a \textit{vanilla} flavor containing
universal platform-independent quantum gates. Using this flavor of the \netqasm
language enables the platform-independent description of quantum network
programs. Platform-\textit{specific} flavors may be created to have quantum
operations that are native and optimized for a specific hardware platform. As an
example, we show a flavor tailored to the Nitrogen-Vacancy (NV) hardware, a
promising platform for quantum network end-nodes~\cite{Taminiau2014,
    hanson2021realization}.

In our model, application-specific classical communication only happens at the
application layer (\cref{fig:app_programs}). In particular, this means that
\netqasm contains no provision for classical communication with the remote node.
We remark that of course, classical control communication may be used by the
\QNPU to realize the services of the quantum network stack accessed through
\netqasm.

With \netqasm\, we solve
various problems that are unique to quantum internet programming:
(1)
for remote
entanglement generation, we introduce new instruction types for making use of an
underlying quantum network stack~\cite{dahlberg2019linklayer, kozlowski2020networklayer},
(2)
for the close interaction between
classical and quantum operations, we use a shared-memory model for sharing
classical data between the application layer and the \QNPU,
(3)
in order to run multiple applications on the same quantum node---which may be beneficial for overall resource usage (see~\cref{sec:design_considerations})---
we make use of virtualized quantum memory, similar to virtual memory in classical computing~\cite{arpaci2018operating},
(4)
since on some platforms, not all qubits may be used to generate remote
entanglement, we introduce the concept of unit-modules describing qubit
topologies with additional information per (virtual) qubit about which operations are
possible.

Since \netqasm\ is meant to be low-level, similar in nature to classical
assembly languages, we have also developed a higher-level software development
kit (SDK), in Python, to make it easier to write applications. This SDK and
related tools are open-source and freely available at~\cite{git_netqasm}, as
part of our Quantum Network Explorer~\cite{qne_website}. Through the SDK we have
also enabled the quantum network simulators
\netsquid~\cite{coopmans2021netsquid} and
\simulaqron~\cite{dahlberg2018simulaqron} to run any application programmed in
\netqasm.

We have evaluated \netqasm\ by simulating the execution of a teleportation
application and a blind quantum computation using \netqasm. Hereby we have shown
that interesting quantum internet applications can indeed be programmed using
\netqasm. Furthermore, the evaluations argue certain design choices of \netqasm,
namely the use of so-called \textit{unit modules}, as well as platform-specific
\textit{flavors}.

We remark that \netqasm has already been used on a real hardware setup in the
lab, in a highly simplified test case that only produces
entanglement~\cite{pompili2021experimental}.

\subsection{Related Work}\label{sec:related} In the field of quantum computing,
a substantial amount of progress has been made related to developing
architectures (e.g.~\cite{fu2017microarchitecture,bourassa2020photonicblueprint,
      murali2019fullstack, wecker2014liqui, khammassi2020openql, amy2019staq,
      green2013quipper, Steiger2016}), instruction sets
(e.g.~\cite{cross2017openqasm,khammassi2018cqasm,fu2019eqasm,liu2017fqasm,smith2016quil,qiskit,cirq,qsharp,jones2019quest})
and compilers~\cite{zulehner2019compiling, haner2018software,
      gokhale2020quantum, liu2020new, gokhale2020optimized, ding2020square,
      smith2020opensource, Sivarajah_2020, hietala2019verified,
      zhang2020contextmapping, niu2020hardware, dury2020qubo, pozzi2020using,
      Nishio_2020}. One example is \qasm, an instruction set framework, borrowing
ideas from classical assembly languages, which has gained a lot of popularity
over the years and has been successfully integrated in software stacks for
quantum computers. There are in fact many variants of \qasm such as
\openqasm~\cite{cross2017openqasm}, \nq{cQASM}~\cite{khammassi2018cqasm},
\nq{eQASM}~\cite{fu2019eqasm}, \nq{f-QASM}~\cite{liu2017fqasm}. Some of these
variants are at a level closer to the physical implementation, such as
\nq{eQASM}, allowing for specifying low-level timing of quantum operations,
while others, such as \nq{f-QASM}, are at a higher level. Together with the
definition of these \qasm-variants, progress has also been made in compilation
of applications programmed in \qasm\ to hardware implementations. More abstract
languages and programming frameworks for quantum programs include
\nq{Quil}~\cite{smith2016quil}, \nq{Qiskit}~\cite{qiskit},
\nq{Cirq}~\cite{cirq}, \nq{Q\#}~\cite{qsharp}, \nq{QuEST}~\cite{jones2019quest}.

None of these instruction sets or languages contain elements for remote
entanglement generation (i.e. between different nodes), which \netqasm does
provide. A \netqasm program that uses the vanilla flavor and only contains
local operations would look similar to an \openqasm program. However, the
instruction set is not quite the same, since \netqasm uses a different memory
model than \openqasm. This is due to the hybrid nature of quantum network
programs, which has more interaction between classical data and quantum data
than non-networking programs (for which \openqasm might be used). So,
\netqasm is not just a superset of the \openqasm instruction set (in the sense of
adding entanglement instructions).

In~\cite{dahlberg2018simulaqron}, we introduced the \cqc interface, which was a
first step towards a universal instruction set. However, \cqc had a number of
drawbacks, in particular: (1) \cqc does not have a notion of virtualized memory
(see \cref{sec:design_considerations}), which meant that applications needed to
use qubit IDs that were explicitly provided by the underlying hardware. This
introduced more communication overhead and fewer optimization opportunities for
the compiler. (2) \cqc does not provide as much information about hardware
details. Therefore, platform-specific compilation and optimization is not
possible. (3) Furthermore, \cqc does not match entirely with the later
definition of our quantum network stack~\cite{dahlberg2019linklayer,
      kozlowski2020networklayer}. For example, it was not clearly defined how \cqc
relates to the definition of a network layer.

Many of the ideas from e.g. \qasm\ for how to handle and compile local gates can
be reused also for quantum network applications. For example, version 3 of
\openqasm~\cite{cross2021openqasm} which is under development, proposes close
integration between \emph{local} classical logic and quantum operations, which
is something we also propose in this work. However, there are two key
differences that we need to address:
\begin{enumerate}
      \item Instructions for generating entanglement between remote nodes in the
            network need to be handled and integrated with the rest of the application,
            see \cref{sec:abstract_model} below.
      \item The local operations performed by a node might depend on information
            communicated by another node and only known at runtime. Note that this
            is different from the conditionals on \emph{local} classical
            information, proposed in for example \openqasm version 3, which does
            not require communication between remote nodes in a network. This brings
            new constraints in how to handle memory allocation, scheduling and
            addressing. We discuss this point in further detail in the coming
            sections.
\end{enumerate}
\netqasm\ solves the above two points and improves upon \cqc.

\subsection{Outline}
In \cref{sec:preliminaries} we define relevant concepts and introduce the
model of end-nodes that we use, including the \QNPU.
In \cref{sec:use_cases} we discuss use-cases of a quantum network which
\netqasm\ should support. In \cref{sec:design_considerations} we consider requirements and
considerations any instruction set architecture for quantum networks should
fulfill which then lay the basis for the decisions that went into developing
\netqasm, see \cref{sec:design_decisions}. In \cref{sec:implementation} and
\cref{sec:python-sdk} we describe details about the \netqasm\ language and
associated SDK. In \cref{sec:evaluation} we quantitatively evaluate some of the
design decision of \netqasm\ by benchmarking quality of execution using the
quantum network simulator \netsquid~\cite{netsquid,coopmans2021netsquid}. We
conclude in \cref{sec:conclusion}.
\section{Preliminaries and Definitions}\label{sec:preliminaries}

\subsection{Quantum networks}\label{sec:quantum_networks} 
\begin{figure}[h]
    \centering
    \includegraphics[width=0.7\textwidth]{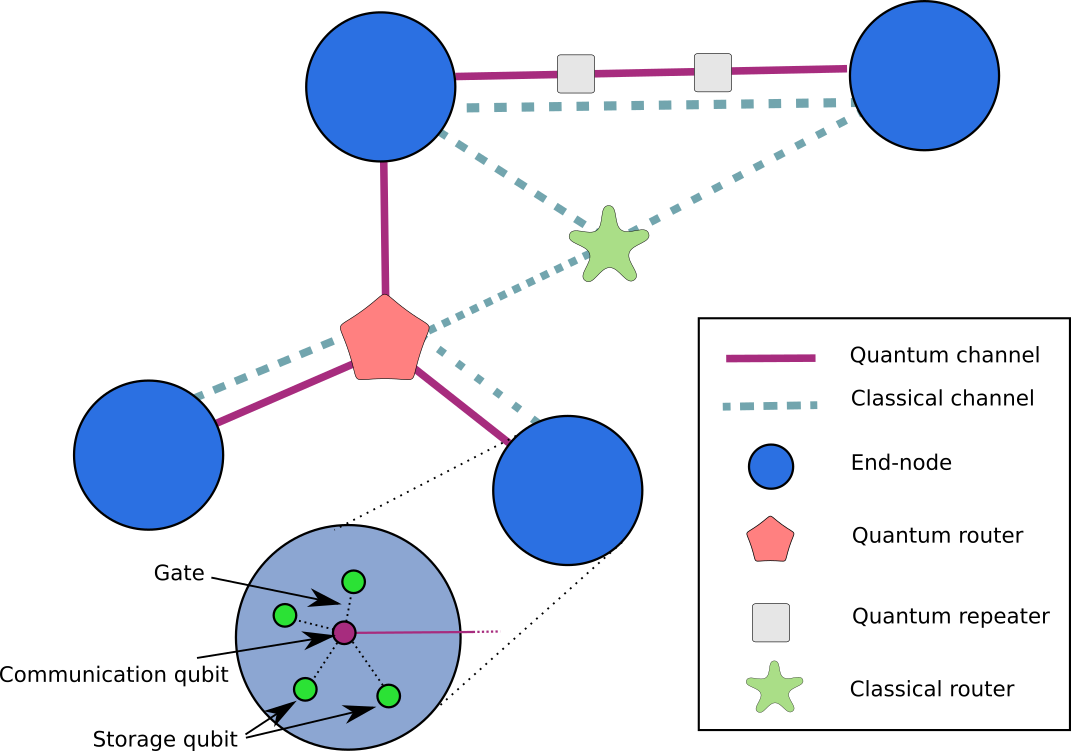}
    \caption{Abstract model of a quantum network and its components. Quantum
        network applications run on the \emph{end-nodes} (blue). Their
        communication via classical message passing and quantum
        entanglement~(\cref{fig:app_programs}) is abstracted away by a network
        stack. That is, it is not visible at the application layer how
        entanglement generation or classical message passing is realized. This
        may be via direct physical connections, or intermediary repeaters and/or
        routers. End-nodes hold two types of qubits: (1) \emph{communication
            qubits} which can be used to generate entanglement with remote nodes and
        (2) \emph{storage qubits} which can be used to store quantum states and
        apply operations. A communication qubit may also be used as a storage
        qubit. The qubits within an end-node can interact through quantum gates
        and their state can be measured.}\label{fig:network_model}
\end{figure}

A schematic overview
of quantum networks is given in~\cref{fig:network_model}. A quantum network
consists of \textit{end-nodes} (hereafter also: \textit{nodes}), which contain
quantum network processors as well as classical processors. Nodes are connected
by \textit{quantum channels} or \textit{links} that can be used to generate
\textit{entangled} quantum states across nodes. End-nodes possess quantum memory
in the form of qubits, which can be manipulated by performing
\textit{operations} such as initialization, readout, and single- or multi-qubit
\textit{gates}. Each quantum memory has a certain \textit{topology} that
describes which operations can be applied on which (pair of) qubits. Some of the
qubits in a quantum memory may be used to generate an entangled state with
another node. These qubits are called \emph{communication
    qubits}~\cite{dahlberg2019linklayer}, in contrast to \emph{storage qubits} which
can only directly interact with other qubits part of the same local
node\footnote{A storage qubit may however hold a state that is entangled with a
    qubit in another node: after remote entanglement generation using a
    communication qubit, the state in that local qubit could be transferred to one of
    the storage qubits, preserving the remote entanglement.}.

Some platforms only have a single communication qubit and multiple storage
qubits~\cite{Bernien2014}, whereas others can have multiple communication
qubits~\cite{Inlek2017}. Qubits are sensitive to \textit{decoherence} and have
limited lifetimes. Therefore, the timing and duration of operations (such as
local gates or entanglement generation with another node) have an impact on the
quality of quantum memory. Classical processors control the quantum hardware,
and also perform classical computation. Finally, classical links exist between
nodes for sending classical messages.

Since end-nodes can control their memory and entanglement generation, they can
run arbitrary \textit{user programs}. End-nodes can both communicate
classically and generate entanglement between each other, either directly or
through repeaters and routers, (\cref{fig:network_model}). Nodes in the network
other than end-nodes, such as repeaters and routers, do not execute
user programs; rather these run protocols that are part of some level in the
network stack~\cite{dahlberg2019linklayer,kozlowski2020networklayer}. These
internal nodes in the network perform elementary link generation and
entanglement swapping in order to generate long-distance remote entanglement
between end-nodes~\cite{dahlberg2019linklayer}.

There are various quantum hardware implementations for quantum network processors,
such as nitrogen-vacancy centers in diamond~\cite{Bernien2014}, ion
traps~\cite{moehring2007entanglement}, and neutral
atoms~\cite{hofmann2012heralded,ritter2012elementary}, which all have different
capabilities and gates that can be performed.

In contrast to classical networks, we consider the end-nodes in a quantum
network to not have a network interface component that is separated
from the main processing unit. Having local and networking operations combined
in a single interface reflects the physical constraint on current and near-term
hardware. Current state-of-the-art hardware for quantum networking devices can
make use of up to the order of 10 qubits~\cite{bradley2019solidstate}.
Furthermore, certain hardware implementations, such as nitrogen-vacancy centers
in diamond~\cite{Bernien2014}, only have a single communication qubit, which
also acts as a mediator for any local gate on the storage qubits. This prevents
dedicating some qubits for purely local operations and some for purely
networking operations. Rather, to make maximal use of near-term quantum
hardware, a multi-purpose approach needs to be supported.

\subsection{Application layer and QNPU}\label{sec:abstract_model}

In this work we will assume an abstract model of the hardware and software
architecture of end-nodes in a quantum network. Specifically, we assume each
end-node to consist of an \host and a \emph{Quantum Network Processing Unit}
(\QNPU). The \host can be also be seen as a the user space of a classical
computer, and the \QNPU as a coprocessor.

This model takes into account both physical- and application-level constraints
found in quantum network programming. The \QNPU\ can be accessed by the \host,
at the same node, to execute quantum and classical instructions. We define the
capabilities of the \QNPU, and roughly their internal components, but do not
assume how exactly this is implemented. In the rest of this work, we simply use
the \QNPU as a black box.

The \QNPU can do both classical and quantum operations, including (1) local
operations such as classical arithmetic and quantum gates and (2) networking
operations, i.e. remote entanglement generation. The \host\ cannot do any
quantum operations. It can only do local computation and classical communication
with other nodes. In terms of classical processing power, the difference between
the \host and the \QNPU is that the \host can do heavy and elaborate
computation, while we assume the \QNPU to be limited in processing power.

The \host\ can interact with the \QNPU\ by for example sending instructions to do
certain operations. The \host\ and the \QNPU\ are logical components and may or may not
be the same physical device. It is assumed that there is low latency in the
communication between these components, and in particular that they are
physically part of the same node in the network.

One crucial difference between the \host and the \QNPU is that the \host can do
application-level classical communication with other end-nodes, while the \QNPU
cannot. The \QNPU can communicate classically to synchronize remote entanglement
generation, but it does not allow arbitrary user-code classical communication.
We use this restriction in order for the \QNPU to have relatively few resource
requirements.

\begin{figure}
    \centering
    \includegraphics[width=0.8\textwidth]{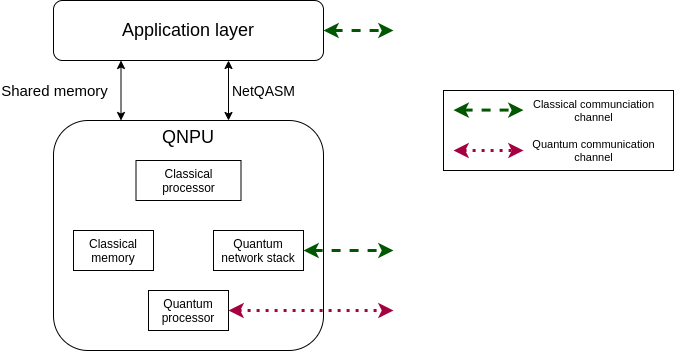}
    \caption{Overview of \QNPU\ components and interfaces. The \host\ talks to
        the \QNPU\ using \netqasm. The processor inside the \QNPU\ can interact with
        all other components. Channels are connecting components with corresponding
        components in adjacent nodes in the network.}
    \label{fig:qnpu}
\end{figure}

The \QNPU\ consists of the following components, see \cref{fig:qnpu}:
\begin{itemize}
    \item \textbf{Processor:} The processor controls the other components of the
          \QNPU\ and understands how to execute the operations specified by the
          \host. It can read and write data to the classical memory and use this
          data to make decisions on what operations to do next. It can apply
          quantum gates to the qubits in the quantum memory and measure them as
          well. Measurement outcomes can be stored in the classical memory.
    \item \textbf{Classical memory:} Random-access memory storing data produced
          during the execution of operations, such as counters, qubit measurement
          outcomes, information about generated entangled pairs, etc.
    \item \textbf{Quantum memory:} Consists of communication and storage qubits,
          see \cref{sec:quantum_networks}, on which quantum gates can be applied.
          The qubits can be measured and the resulting outcome stored in the
          classical memory by the processor. The communication qubits are
          connected through a quantum channel to adjacent nodes in the quantum
          network, through which they can be entangled. This quantum channel may
          also include classical communication needed for synchronization, phase
          stabilization or other mechanisms needed in the specific realization.
    \item \textbf{Quantum network stack:} Communicates classically with other
          nodes and quantum repeaters in the network to synchronize the
          generation of remote entanglement, and issues low-level instructions
          to execute the entanglement generation procedures,
          see~\cite{dahlberg2019linklayer,kozlowski2020networklayer}.
\end{itemize}

We stress that the internals of the \QNPU are not relevant to the design of
\netqasm. We do assume that the \QNPU only has limited classical processing
power, and can therefore be implemented on for example a simple hardware board.

\subsection{Applications and programs}
As mentioned in~\cref{sec:introduction}, quantum network \textit{applications}
(or protocols) are multi-partite and distributed over multiple end-nodes. The
unit of code that is executed on each of the end-nodes that are part of the
application, is called a \textit{program}. We will use this terminology
throughout the rest of the paper.

As mentioned in the previous section, the end-nodes are modeled such that there
is an application layer and a \QNPU. We assume that execution of quantum network
programs is handled by the application layer. How exactly the program is
executed, and how the \QNPU is involved herein, is part of the \netqasm
proposal.

\section{Use-cases}\label{sec:use_cases} In the next section we will discuss the
design considerations taken when developing \netqasm. These design
considerations are based on a set of use-cases listed in this section which we
intend for \netqasm\ to support. Applications intended to run on a quantum
network will often depend on a combination of these use-cases.

\begin{itemize}
      \item \textbf{Local quantum operations}. Applications running on a
            network node need to perform quantum operations on local qubits,
            including initialization, measurement, and single- or multi-qubit
            gates. Such local qubit manipulation is well known in the field of
            quantum computing. For example, \openqasm~\cite{cross2017openqasm}
            describes quantum operations. Quantum \textit{network}
            applications should be able to do these local operations as well.

      \item \textbf{Local quantum operations depending on local events or data}.
            The next use-case stems from applications consisting of
            programs in which limited classical computation or decision
            making is needed in-between performing quantum operations.
            Here we consider only dependencies in a program between
            quantum operations and information that is produced locally,
            that is, on the node that this program is being executed. For
            instance, a program might only apply a quantum gate on a qubit
            depending on the measurement outcome of another qubit, or
            choose between execution branches based on evaluation of a
            classical function of earlier measurement outcomes. An example
            is for the server-side of \textit{blind quantum computation},
            which performs a form of Measurement-Based Quantum Computation
            (MBQC). In each step of the MBQC, the server performs certain
            gates on a qubit, depending on results of measuring previous
            qubits~\cite{fitzsimons2017private}. These applications need
            classical operations to not take too much time, so that qubit
            states stay coherent during these operations. This implies
            that switching between classical and quantum operations should
            have little overhead.

      \item \textbf{Entanglement generation}. Crucial to quantum networks
            is the ability to generate remote \textit{entanglement}.
            Applications should be able to specify requests for entanglement
            generation between remote nodes. In some cases, a Measure-Directly
            (MD)~\cite{dahlberg2019linklayer} type generation is required, where
            entangled state is measured directly, without storing in memory, to
            obtain correlated classical bits, such as in Quantum Key
            Distribution (QKD). However, in many cases a Create-Keep
            (CK)~\cite{dahlberg2019linklayer} type is needed, where the
            entanglement needs to be stored in memory and further operations
            applied involving other qubits. We want applications to be able to
            \textit{initiate} or \textit{receive} (await) entanglement of both
            forms with nodes in the network.

      \item \textbf{Local quantum operations depending on remote events or
                  data}. We already mentioned the use-case of having
            conditionals based on \emph{local} information. We also
            envision applications that need to store qubits and
            subsequently perform local quantum operations on them and
            other local qubits, based on classical information coming from
            \textit{another node}. An example is \textit{teleportation} in
            which the receiver---after successful entanglement
            generation---needs to apply local quantum corrections based on
            the measurement outcomes of the sender. Another application is
            blind quantum computation, where the server waits for
            classical instructions from the client about which quantum
            operations to perform. Hence, there need to be integration of
            classical communication (sending the measurement results or
            further instructions) and the local quantum operations.
            Furthermore, since classical communication has a non-zero
            latency (and is in general even non-deterministic), it should
            be possible to suspend doing quantum operations while waiting
            for communication or performing classical processing, while
            quantum states stay coherent.

      \item \textbf{Waiting time}. We consider the scenario where an
            application requires two nodes to communicate with each other, and
            where communication takes a long time, for example since they are
            physically far apart. It should be possible for a program to suspend
            doing quantum operations while waiting for communication or
            performing classical processing, while quantum states stay coherent.
            Furthermore, in order to maximize the usage of the \QNPU we want to
            have a way to fill this waiting time in a useful way.
\end{itemize}

\section{Design Considerations}\label{sec:design_considerations} In this section we review the
most important design considerations and requirements that were applied when
developing \netqasm. Our proposed solutions to these design considerations are
presented in the next section, with more details about \netqasm\ as a language
in the subsequent sections.

\begin{itemize}
      \item \label{item:design_ent_gen} \textbf{Remote entanglement generation:}
            One of the main differences compared to the design considerations of
            a quantum computing architecture is that of remote entanglement
            generation (see the use-case in~\cref{sec:use_cases}). Nodes need to
            be able to generate entanglement with a remote node, which requires
            the collaboration and synchronization of both nodes, and possibly
            intermediate nodes, which is handled by the network
            stack (\cref{sec:preliminaries}).

            Further requirements arise in platforms with a limited number of
            communication qubits. The extreme case is nitrogen-vacancy centers
            in diamond which have a single communication qubit that additionally
            is required for performing local operations. For this reason it is
            not possible to decouple local gates on qubits from entanglement
            generation. We note the contrast with classical processors, where
            networking operations are typically intrinsically separate kinds
            of operations. For example, operations such as sending a message may
            simply involve moving data to a certain memory (e.g. that of a
            physically separate network interface), which is often abstracted as
            a system call.

            A quantum network stack has already been proposed
            in~\cite{dahlberg2019linklayer,kozlowski2020networklayer}, and we
            expect the \QNPU of the end-node to implement such a stack,
            including a \textit{network layer} that exposes an interface for
            establishing entanglement with remote nodes. The way in which a
            program creates remote entanglement should therefore be compatible
            with this network layer.

      \item \label{item:design_cond} \textbf{Conditionals:}
            In~\cref{sec:use_cases} we mentioned the need to do local quantum
            operations conditioned on classical data that may be generated locally or
            by remote nodes. Such classical data include for example measurement
            results or information communicated to or from other nodes in the network.
            We distinguish between real-time and near-time
            conditionals~\cite{cross2021openqasm}. Real-time conditionals are
            time-sensitive, such as applying a certain quantum operation on a qubit
            depending on a measurement outcome. For such conditionals, we would like
            to have fast feedback, in order for quantum memory not to wait too long
            (which would decrease their quality). Near-time conditionals are not as
            sensitive to timing. For example, a program may have to wait for a
            classical message of a remote node, while no quantum memory is currently
            being used. Although it is preferably minimized, the actual waiting time
            does not affect the overall execution quality.

      \item \label{item:design_return} \textbf{Shared memory:} As described in
            \cref{sec:preliminaries}, we expect end-nodes to consist of an
            application layer and a \QNPU. These two components have different
            capabilities. For example, only the application layer has the
            ability to do arbitrary classical communication with other nodes.
            Only the \QNPU can do quantum operations. These restrictions lead
            the design in a certain way.
            The two components hence need to work together somehow.
            There needs to be model for interaction between the two, and also
            for shared memory.

            Executing programs on an end-node is shared by the application layer
            and the \QNPU (see~\cref{sec:abstract_model}). Indeed, only the
            \QNPU can do quantum-related operations, whereas the application
            layer needs to do classical communication. In order to make these
            work together, the two components have to share data somehow. This
            includes the application layer requesting operations on the \QNPU,
            and sending the following from the \QNPU to the \host: (1)
            measurement outcomes of qubits, (2) information about entanglement
            generation, in particular a way to identify entangled pairs. This
            communication between \host and \QNPU needs to be done during
            runtime of the program. This is in contrast to local quantum
            computation, where one might wait until execution on the \QNPU is
            finished before returning all data. The challenge for quantum
            network programs is to have a way to return data while quantum
            memory stays in memory.

      \item \label{item:design_proc_delay} \textbf{Processing delay:} Since we
            assume that the application layer and the \QNPU have to share
            execution of a single program, the interaction between the two
            layers should be efficient. Unnecessary delays lead to reduced
            quality (see~\cref{sec:introduction}). The challenge is therefore to
            come up with an architecture for the interaction between the
            application layer and the \QNPU, as well as a way to let \QNPU
            execution not take too long.
      \item \label{item:design_pi} \textbf{Platform-independence:}
            As explained in~\cref{sec:introduction}, hardware can have many
            different capabilities and gates that can be performed. However,
            application programmers should not need to know the details of the
            underlying hardware. For this reason, there needs to be a framework
            through which a programmer can develop an application in a
            platform-independent way which compiles to operations the \QNPU\ can
            execute.
      \item \label{item:design_opt} \textbf{Potential for optimization:} Since near-term
            quantum hardware has a limited number of qubits and qubits have a
            relatively short lifetime, the hardware should be utilized in an
            effective way. There is therefore a need to optimize the quantum
            gates to be applied to the qubits. This includes for example
            choosing how to decompose a generic gate into native gates,
            rearranging the order of gates and measurements and choosing what
            gates to run in parallel. Since different platforms have vastly
            different topologies and gates that they can perform, this
            optimization needs to take the underlying platform into account.
            The challenge is to have a uniform way to express both platform-independent
            and platform-specific instructions.
      \item \label{item:design_parallel} \textbf{Multitasking:} The `Waiting
            time' use-case in~\cref{sec:use_cases} describes that a node's \QNPU
            may have to wait a long time. We consider the solution that the
            \QNPU may do multitasking, that is, run multiple (unrelated)
            programs at the same time. Then, when one program is waiting,
            another program can execute (partly) and fill the gap. To make our
            design compatible with such multitasking, we need to provide a
            way such that programs can run at the same time as other programs,
            but without having to know about them.
      \item \label{item:ease_programming} \textbf{Ease of programming:} Even
            though \netqasm\ provides an abstraction over the interaction with the
            \QNPU, it is still low-level and hence not intended to be used
            directly by application developers. Furthermore, applications also
            contain classical code that is not intended to run on the \QNPU.
            Therefore it should be possible to write programs consisting of both
            classical and quantum (network) operations in a high-level language
            like Python, and compile them to a hybrid quantum-classical program
            that uses \netqasm.

\end{itemize}

\section{Design Decisions}\label{sec:design_decisions} Based on the use-cases, design
considerations and requirements, we have designed the low-level language
\netqasm\ as an API to the \QNPU. In this section we present concepts and design
decisions we have taken. Details on the mode of execution and the
\netqasm-language are presented in \cref{sec:implementation}.

\subsection{Interface between \host and QNPU}\label{sec:design_decisions_interface}
\subsubsection{Execution model}
As described in \cref{sec:preliminaries}, and also in \cref{sec:design_considerations} program
execution is split across the \host and the \QNPU.
Since the \QNPU is assumed to have limited processing power
(\cref{sec:preliminaries}), our design lets the \host do most of the
classical processing. The program blocks (\cref{fig:program_decomp}) are hence
spread over two separate systems: blocks of purely classical code are executed
by the \host, and blocks of quantum code (containing both quantum operations and limited
classical control) are executed by the \QNPU.

The quantum code (including limited classical control) is expressed using the
\netqasm language. The classical code is handled completely by the \host, and we
do not impose a restriction to its format. In our implementation
(\cref{sec:python-sdk}), we use Python. This classical code on the \host
also handles all application-level classical communication between nodes, since
it cannot be done on the \QNPU.

We let the \host initiate a program. Whenever quantum code needs to be executed,
the \host delegates this to the \QNPU. Since processing delay should be
minimized (\cref{sec:design_considerations}), the communication between \host
and \QNPU should be minimized. Therefore, \netqasm bundles the quantum
operations together into blocks of instructions, called \textit{subroutines}, to
be executed on the \QNPU. A program, then, consists of both both classical code
and quantum code, and the quantum code is represented as one or more
subroutines. These subroutines can be seen as the quantum code blocks
of~\cref{fig:program_decomp}.

For most programs, we consider subroutines to be sent
consecutively in time. However, if the \QNPU supports it, \netqasm also allows
to send multiple subroutines to be executed on the \QNPU at the same time,
although this requires some extra care when dealing with shared memory. From the
perspective of the \QNPU, a program consists of a series of subroutines
sent from the \host. Before receiving subroutines, the \host first
\textit{registers} a program at the \QNPU. The \QNPU then sets up the
classical and quantum memories (see below) for this program. Then, the \host
may send subroutines to the \QNPU for execution.

\subsubsection{Shared classical memory}
Since classical and quantum blocks in the code (as per
\cref{fig:program_decomp}) can depend on each other, the \host and the \QNPU
need to have a way to communicate information to each other.
For example, a subroutine may
include a measurement instruction; the outcome of this measurement may be used
by the \host upon completion of the subroutine.
Therefore, \netqasm uses a shared memory model such that conceptually both
layers can access and manipulate the same data. This solves the need to return
data, and to do conditionals (\cref{sec:design_considerations}).

Each program has a classical memory space consisting of \textit{registers} and
\textit{arrays}. Registers are the default way of storing classical values, like
a measurement outcome. In the example of the \host needing a measurement
outcome, there would be an instruction in the subroutine saying that a
measurement outcome needs to be placed in a certain register. The \host can then
access this same register (since they share the memory space) and use it in
further processing. The number of registers is small, and constant for each
program. Arrays are collections of memory slots (typically the slots are
contiguous), which can be allocated by the program at runtime. Arrays are used
to store larger chunks of data, such as parameters for entanglement requests,
entanglement generation results, or multiple measurement outcomes when doing
multiple entangle-and-measure operations. The \host may only read from the
shared memory; writing to it can only be done by issuing \netqasm instructions
such as \texttt{set} (for registers) and \texttt{store} (for arrays). The \QNPU
may directly write to the shared memory, for example when entanglement finished
and it writes the results to the array specified by the program.

\begin{figure}
      \centering
      \begin{subfigure}[b]{\textwidth}
            \centering
            \begin{tikzpicture}[
                  host-col/.style={red, thick},
                  qdevice-col/.style={black!40!green, thick},
                  arrow/.style={black, ->, thick},
                  process/.style={black, Square[]-Square[], thick},
                  arrow-label/.style={right, near start},
                  ]
                  \def\height{\textwidth/6};
                  \def\width{0.8\textwidth}
                  \coordinate (upleft) at (-\width/2,\height/2);
                  \coordinate (upright) at (\width/2,\height/2);
                  \coordinate (downleft) at (-\width/2,-\height/2);
                  \coordinate (downright) at (\width/2,-\height/2);

                  \draw [host-col] (upleft) -- (upright);
                  \draw [qdevice-col] (downleft) -- (downright);

                  \def\x{7}  
                  \draw [arrow] (upleft) -- node[arrow-label] {Circuit (e.g QASM)} ($(downleft)!1/\x!(downright)$);
                  \draw [arrow] ($(upleft)!4/\x!(upright)$) -- node[arrow-label] {Circuit (e.g. QASM)} ($(downleft)!5/\x!(downright)$);
                  \draw [arrow] ($(downleft)!2/\x!(downright)$) -- node[arrow-label] {Outcome} ($(upleft)!3/\x!(upright)$);
                  \draw [arrow] ($(downleft)!6/\x!(downright)$) -- node[arrow-label] {Outcome} ($(upleft)!7/\x!(upright)$);

                  \def\d{0.2}
                  \draw [process] ($(downleft)!1/\x!(downright) + (0,-\d)$) -- node[below] {Execution} ($(downleft)!2/\x!(downright) + (0,-\d)$);
                  \draw [process] ($(downleft)!5/\x!(downright) + (0,-\d)$) -- ($(downleft)!6/\x!(downright) + (0,-\d)$);
                  \draw [process, black!5!orange] ($(downleft)!2/\x!(downright) + (\d,-\d)$) -- node[below] {Reset qubits} ($(downleft)!5/\x!(downright) + (-\d,-\d)$);
                  \draw [process] ($(upleft)!3/\x!(upright) + (0,\d)$) -- node[above, align=center] {Classical\\comp.} ($(upleft)!4/\x!(upright) + (0,\d)$);

                  \node (host) [host-col, left=of upleft, xshift=20pt] {Application layer};
                  \node (qdevice) [qdevice-col, left=of downleft, xshift=20pt, align=center] {Quantum\\computing\\device};
            \end{tikzpicture}
            \caption{}\label{fig:hybrid:a}
      \end{subfigure}\\
      \begin{subfigure}[b]{\textwidth}
            \centering
            \begin{tikzpicture}[
                  host-col/.style={red, thick},
                  qdevice-col/.style={black!40!green, thick},
                  arrow/.style={black, ->, thick},
                  process/.style={black, Square[]-Square[], thick},
                  arrow-label/.style={right, near start},
                  ]
                  \def\height{\textwidth/6};
                  \def\width{0.8\textwidth}
                  \coordinate (upleft) at (-\width/2,\height/2);
                  \coordinate (upright) at (\width/2,\height/2);
                  \coordinate (downleft) at (-\width/2,-\height/2);
                  \coordinate (downright) at (\width/2,-\height/2);

                  \draw [host-col] (upleft) -- (upright);
                  \draw [qdevice-col] (downleft) -- (downright);

                  \def\x{7}  
                  \draw [arrow] (upleft) -- node[arrow-label] {Quantum code block} ($(downleft)!1/\x!(downright)$);
                  \draw [arrow] ($(upleft)!4/\x!(upright)$) -- node[arrow-label] {Quantum code block} ($(downleft)!5/\x!(downright)$);
                  \draw [arrow] ($(downleft)!2/\x!(downright)$) -- node[arrow-label] {Outcome} ($(upleft)!3/\x!(upright)$);
                  \draw [arrow] ($(downleft)!6/\x!(downright)$) -- node[arrow-label] {Outcome} ($(upleft)!7/\x!(upright)$);

                  \def\d{0.2}
                  \draw [process] ($(downleft)!1/\x!(downright) + (0,-\d)$) -- node[below] {Execution} ($(downleft)!2/\x!(downright) + (0,-\d)$);
                  \draw [process] ($(downleft)!5/\x!(downright) + (0,-\d)$) -- ($(downleft)!6/\x!(downright) + (0,-\d)$);
                  \draw [process, black!5!blue] ($(downleft)!2/\x!(downright) + (\d,-\d)$) -- node[below] {Qubits persists} ($(downleft)!5/\x!(downright) + (-\d,-\d)$);
                  \draw [process] ($(upleft)!3/\x!(upright) + (0,\d)$) -- node[above, align=center] {Classical\\comp.} ($(upleft)!4/\x!(upright) + (0,\d)$);

                  \node (host) [host-col, left=of upleft, xshift=20pt] {Application layer};
                  \node (qdevice) [qdevice-col, left=of downleft, xshift=20pt, align=center] {QNPU};
            \end{tikzpicture}
            \caption{}\label{fig:hybrid:b}
      \end{subfigure}
      \caption{Program interaction between the \host and a quantum device in
            both the case of \emph{hybrid-quantum computing} (\cref{fig:hybrid:a})
            and quantum networks (\cref{fig:hybrid:b}). In the case of hybrid-quantum
            computing, qubits are reset in between circuits (in e.g. \qasm). For
            quantum internet programs the qubits should on the other hand be
            kept in memory, since they might be entangled with another node and
            intended to be used further.}\label{fig:hybrid}
\end{figure}
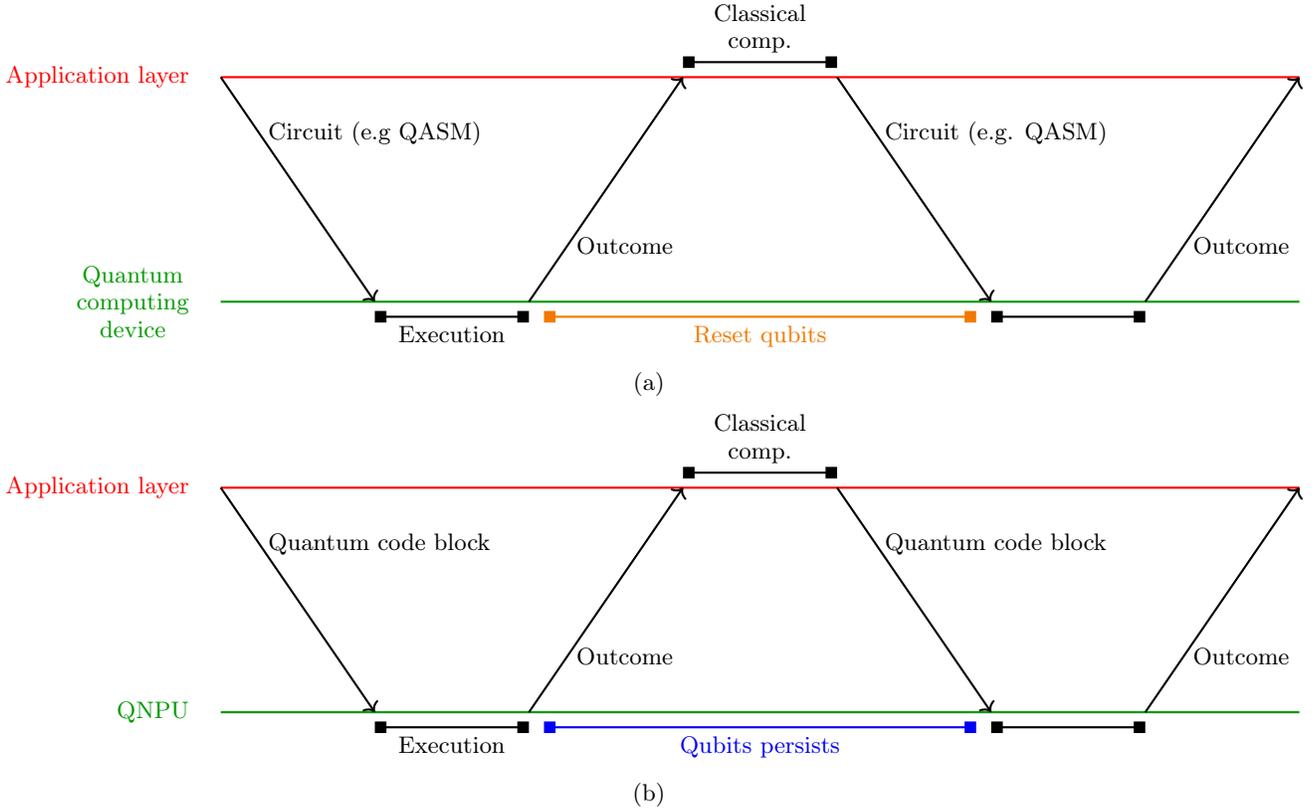

\subsubsection{Unit modules}
In order to support systems with multitasking
(\cref{sec:design_considerations}), \netqasm provides a virtualized model of the
quantum memory to the program. This allows the \QNPU to do mapping between the
virtualized memory and the physical memory and perform scheduling between
programs.

\begin{figure}
      \centering
      \includegraphics[width=0.7\textwidth]{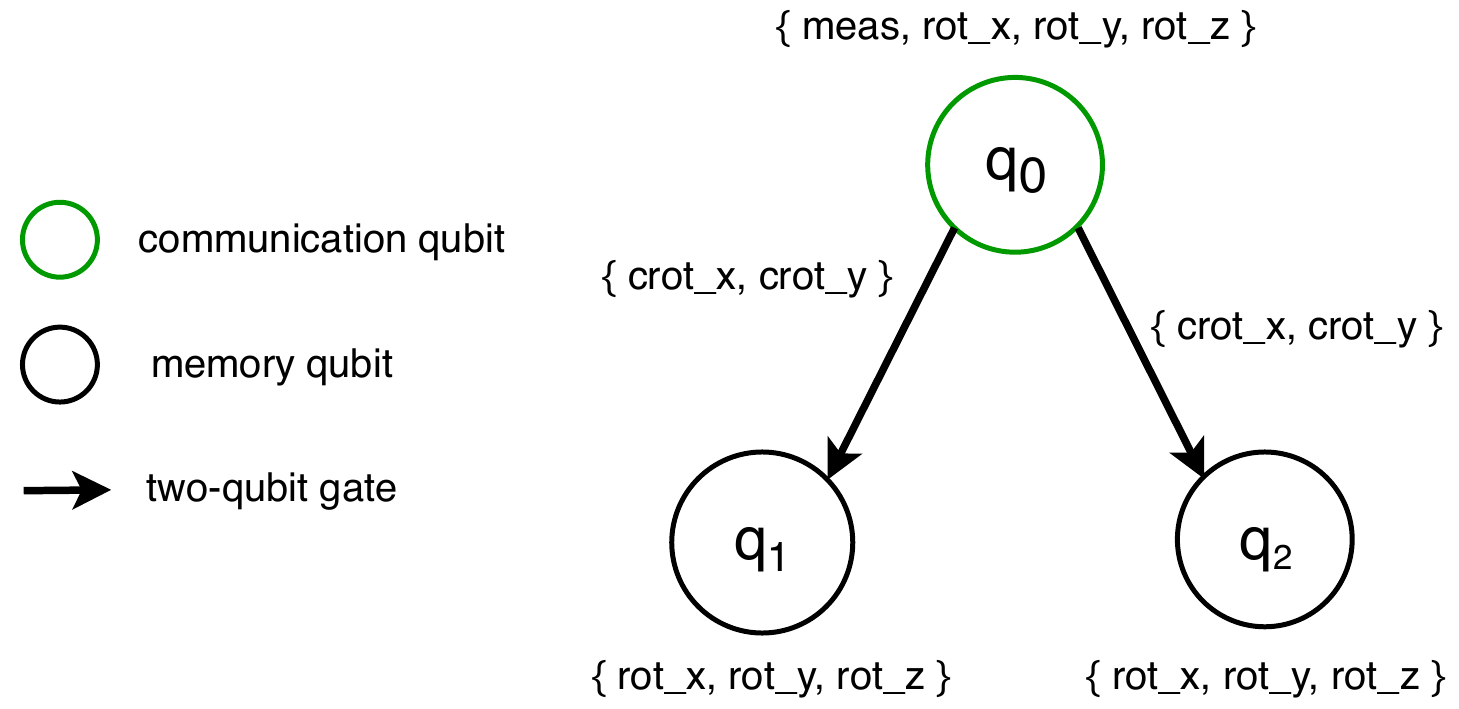}
      \caption{Example of a unit-module topology on a platform using
            nitrogen-vacancy centers in diamond. A unit-module is a
            hypergraph~\cite{berge1984hypergraphs}, with associated information
            on both nodes and edges. Each node represents a virtual qubit,
            containing information about (1) its qubit type (communication or
            storage), (2) physical properties of the qubit, such as decoherence
            times and (3) which single-qubit gates are supported on the qubit,
            together with their duration and noise. Each edge represents the
            possibility of performing joint operations on those qubits, such as
            two-qubit gates, and also containing information about gate
            durations and noise.}\label{fig:topology}
\end{figure}

The quantum memory for a program is represented by a \textit{unit module}
(\cref{fig:topology}). A unit module defines the topology of the available
qubits (which qubits are connected, i.e. on which qubit pairs a two-qubit gate
can be executed), plus additional information on each qubit. This additional
information consists of which gates are possible on which qubit or qubit pair.
It also specifies if a qubit can be used for remote entanglement generation or
not. The extra information is needed since on some platforms, not all qubits can
be used for entanglement generation and different qubits may support different
local gates. For example, in a single NV-centre, there is only one communication
qubit and any additional qubits are storage qubits. Also, the communication
qubit can do different local gates than the storage qubits.

A single program has a single quantum memory space, which is \textit{not} reset
at the end of a subroutine, which is in contrast with quantum computing. This
allows the \host to do processing while qubits are in memory. The following
sequence of operations provides an example. (1) The \host first sends a
subroutine containing instructions for entanglement generation with a remote
node R. (2) The \QNPU has finished executing the subroutine, and informs the
\host about it. There is now a qubit in the program's memory that is entangled
with some qubit in R. (3) The \host does some classical processing and waits for
a classical message from (the \host of) R. (4) Based on the contents of the
message, the \host sends a new subroutine to the \QNPU containing instructions
to do certain operations on the entangled qubit. The subroutine can indeed
access this qubit by using the same identifier as the first subroutine, since
the quantum memory is still the same. We note the contrast with (non-network)
quantum computing, where quantum memory is reset at the end of each block of
instructions (\cref{fig:hybrid}).

Unit modules contain \textit{virtual} qubits. This is because of the requirement
that it should be possible to run multiple programs at the same time on a single
\QNPU (\cref{sec:design_considerations}). Qubits in the unit module are
identified by a \textit{virtual ID}. The \QNPU maps virtual IDs to physical
qubits. A program hence uses a virtual memory space (the unit module), and does not
have to know about the physical memory.

\subsection{NetQASM language}
\subsubsection{Instructions}\label{sec:design_decisions_language} As explained
in~\cref{sec:design_decisions_interface}, the \host delegates quantum code
(including limited classical control) of the program to the \QNPU by creating
blocks of instructions and sending these to the \QNPU for execution. These
blocks are called subroutines and contain \netqasm \textit{instructions}. Since
the \QNPU is meant to be limited in processing power, the instruction set that
it interprets should also be simple and low-level. The \netqasm instruction set
contains instructions for simple arithemetic, classical data manipulation, and
simple control flow in the form of (un)conditional branch instructions. Although
conditional control-flow can be done at the \host as well, \netqasm branching
instructions allow for much faster feedback since they are executed by the
\QNPU, and hence cover the design consideration of real-time conditionals
(\cref{sec:design_considerations}). There are no higher-level concepts such as
functions or for-loops, which would require more complicated and
resource-demanding parsing for the \QNPU, such as constructing an abstract
syntax tree.

A single instruction specifies an operation, possibily acting on classical or
quantum data. For example, a single-qubit rotation gate is represented as an
instruction containing the type of gate, the classical register containing the
rotation angle, and the classical register containing the virtual ID of the
qubit (as specified in the unit module) to act on. \netqasm specifies a set of
\textit{core} instructions that are expected to be implemented by any \QNPU.
These include classical instructions like storing and loading classical data,
branching, and simple arithmetic. Different hardware platforms support different
quantum operations. \netqasm should also support platform-specific optimization
(\cref{sec:design_considerations}). Therefore, \netqasm uses \textit{flavors}
of quantum instructions (\cref{sec:design_decisions_flavours}). The
\textit{vanilla} flavor consists universal of a set of platform-independent
quantum gates. Particular hardware platforms, such as the NV-centre, may use a
special NV flavor, containing NV-specific instructions. A \QNPU implementation
may use a custom mapping from vanilla instructions to platform-specific ones.
The instructions in a flavor are also called a software-visible gate
set~\cite{murali2019fullstack}.
See~\cref{app:instructions} for more details on \netqasm instructions.

\subsubsection{Remote entanglement generation}
Generating entanglement with a remote node is also specified by instructions.
These are however somewhat special compared to other instructions. First,
entanglement generation has a non-deterministic duration. Therefore, when an
entanglement instruction is executed, the request is forwarded to the part of
the system repsonsible for creating entanglement, but the instruction itself
immediately returns. A separate \textit{wait} instruction can be used to block
on entanglement generation to actually be completed. Second, entanglement
generation requests should be compatible with the network stack proposed
in~\cite{dahlberg2019linklayer}, including the network layer
from~\cite{kozlowski2020networklayer}. These requests need to be accompanied by
information such as the number of EPR pairs to generate or the minimum required
fidelity. Third, this information should be able to depend on runtime
information. For example, the required fidelity may depend on an earlier
measurement outcome. Therefore, entanglement generation parameters cannot be
static data, and must be stored in arrays. Furthermore, the result of
entanglement generation with the remote node consists of a lot of information,
such as which Bell state was produced, the time it took, and the measurement
results in case of measuring directly. This information is written by the \QNPU
to an array which is specified by the entanglement instruction. Finally, since
writing the information to the array indicates that entanglement generation
succeeded, the wait instruction can be used to wait until a certain array is
filled in, such as the one provided by the entanglement instruction. Since the
entanglement instruction is non-blocking, it is possible to continue doing local
operations while waiting for entanglement generation to complete.

We assume that the \QNPU implements a network stack where connections need to be
set-up between remote nodes before entanglement generation can
happen~\cite{kozlowski2020networklayer, dahlberg2019linklayer}. \netqasm
provides a way for programs to open such connections in the form of
\textit{EPR sockets}. The \host can ask the \QNPU to open an EPR socket with a
particular remote node. The \QNPU is expected to set up the required connections
in the network stack, and associates this program socket with the
connection. When the program issues an instruction for generating
entanglement, it refers to the EPR socket it wants to use. Based on this, the
\QNPU can use the corresponding connection in the network.

\begin{figure}
      \centering
      \includegraphics[width=0.7\textwidth]{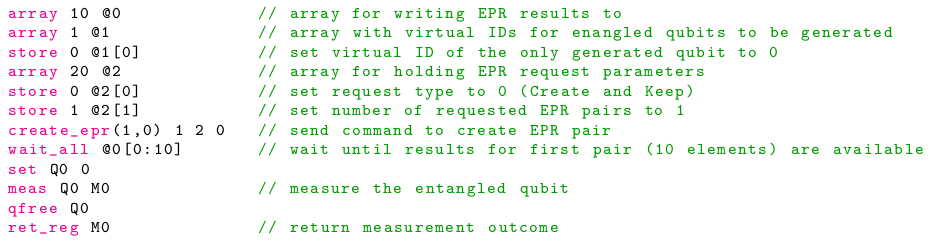}
      \caption{Example of NetQASM code for generating a single entangled pair with anoter node followed by a measurement.
            See the Appendix for more details of the instructions.}
      \label{fig:nqasm_code_example}
\end{figure}

\subsubsection{Flavors}\label{sec:design_decisions_flavours} We want to keep
\netqasm platform-independent. However, we also want the potential for
platform-specific optimization (\cref{sec:design_considerations}). Therefore we
introduce the concept of \textit{flavors}. Flavors only affect the quantum
instruction set of the language, and not the memory model or the interaction
with the \QNPU. We use the \textit{vanilla} or generic flavor for a general,
universal gate set. Subroutines may be written or generated in this vanilla
flavor. Platform-independent optimization may be done on this level. A \QNPU
may directly support executing vanilla-flavored \netqasm. Platform-specific
translations may then be done by the \QNPU itself. It can also be that a \QNPU
only supports a specific flavor of \netqasm. A reason for this could be that
the \QNPU does not want to spend time translating of the instructions at
runtime. In this case, the \host should perfrom a translation step from the
vanilla flavor to the platform-specific flavor. In such a case, the vanilla
flavor can be seen as an \textit{intermediate represenation}, and the
translation to a specific flavor as a back-end compilation step.

\subsubsection{Programmability}
Since the \netqasm instructions are relatively low-level, we like to have a
higher-level programming language for writing programs, that is
automatically compiled to \netqasm. We introduce a higher-level SDK in
\cref{sec:python-sdk}. However, we do not see this as part of the \netqasm
specification itself. This decoupling allows the development of SDKs to be
independent such that these can be provided in various languages and frameworks.

We still want \netqasm instructions to be suitable for manual writing and
inspection. Therefore, instructions (and subroutines) have two formats: a binary
one that is used when sending to the \QNPU, and a text format that is
human-readable. The text format resembles assembly languages including \openqasm.
Example are given in \cref{sec:sdk} and the Appendix.

\section{Implementation}\label{sec:implementation}

\subsection{Interface between \host and \QNPU}
Here we explain the flow of messages between the \host\ and the
\QNPU. The \host\ starts by declaring the registration of an application,
including resource-requirements for the application. After this, the \host\
sends some number of subroutines for the \QNPU\ to execute before declaring the
application is finished. See \cref{fig:message_sequence} for a sequence diagram
and below for a definition of the messages. In \cref{sec:language}
we will describe in more details the content of the subroutines and the format
of instructions. The \QNPU\ returns to the \host\ an assigned application ID for
the registered application and returns data based on the subroutines executed.

The \host and the \QNPU are assumed to run independently and in parallel. For
example, while a subroutine is being executed by the \QNPU, the \host could in
principle do other operations, such as heavy processing or communication with
another node.

\begin{figure}
      \centering
      \begin{sequencediagram}
            \newthread[red]{H}{Application layer}
            \tikzstyle{inststyle}+=[below right=-0.85cm and 2cm of H]
            \newthread[yellow]{Q}{QNPU}
            \mess{H}{RegisterApp}{Q}
            \mess{Q}{RegisterAppOK}{H}
            \mess{H}{Subroutine}{Q}
            \begin{call}{Q}{Execute subroutine}{Q}{}
            \end{call}
            \mess{Q}{Done}{H}
            \mess{H}{Subroutine}{Q}
            \begin{call}{Q}{Execute subroutine}{Q}{}
                  \begin{messcall}{Q}{Update memory}{H}
                  \end{messcall}
            \end{call}
            \mess{Q}{Done}{H}
            \mess{H}{StopApp}{Q}
      \end{sequencediagram}
      \caption{Flow of messages between the \host and the \QNPU.}
      \label{fig:message_sequence}
\end{figure}

\Cref{fig:message_sequence} shows an
example of a message exchange between the \host\ and the \QNPU. The content
of these messages is further detailed in~\cref{sec:app-messages}.

\subsection{The language}\label{sec:language} The syntax and structure of
\netqasm\ resemble that of classical assembly languages, which in turn inspired
the various QASM-variants for quantum computing~\cite{cross2017openqasm,
      khammassi2018cqasm, fu2019eqasm, liu2017fqasm}.

A \netqasm\-instruction is formed
by an instruction name followed by some number of operands:
\begin{nqcode}
      instr operands
\end{nqcode}
where \nq{instr} specifies the instruction, for example \nq{add} to add numbers
or \nq{h} to perform a Hadamard. The \nq{operands} part consists of zero or more
values that specify additional information about the instruction, such as which
qubit to act on in the case of a gate instruction. Instructions and operands are
further specified in~\cref{sec:operands}.

\subsection{Instructions}\label{sec:instructions} 
\begin{figure}
      \centering
      \begin{tikzpicture}[node distance=12pt,
                  flavour-box/.style={rectangle, rounded corners, fill=#1!30, align=center, yshift=-1.8cm,
                              draw=black,
                              minimum height=3cm, minimum width=(\x2-\x1-\pgflinewidth-12pt)/4},
                  flavour-box/.default=green,
                  group-box/.style={rectangle, rounded corners, minimum width=3cm, minimum height=1cm, text centered, draw=black, align=center},
            ]
            \node (classical) [group-box, fill=Cyan] {\textbf{Classical:}\\add, sub\\addm, subm};
            \node (branch) [group-box, right=of classical, fill=LimeGreen] {\textbf{Branch:}\\jmp, bez, bnz,\\beq, bne, blt, bge};
            \node (memory) [group-box, right=of branch, fill=BlueGreen] {\textbf{Memory:}\\set, store, load,\\undef, lea};
            \node (allocate) [group-box, right=of memory, fill=Melon] {\textbf{Allocate:}\\array, qalloc, qfree};
            \node (wait) [group-box, below=of classical, fill=SpringGreen] {\textbf{Wait:}\\wait\_all, wait\_any,\\wait\_single};
            \node (return) [group-box, right=of wait, fill=Orange] {\textbf{Return:}\\ret\_reg, ret\_arr};
            \node (measurement) [group-box, right=of return, fill=Lavender] {\textbf{Measurement:}\\meas};
            \node (entanglement) [group-box, right=of measurement, fill=Orchid] {\textbf{Entanglement:}\\create\_epr, recv\_epr};

            \node (core) [rectangle, above=of $(branch)!0.5!(memory)$, yshift=20pt] {\textbf{Core}};
            \begin{scope}[on background layer]
                  \node (core-bg) [fill=black!10, draw=black, rounded corners, inner sep=12pt,
                        fit={(core) (classical) (branch) (memory)
                                    (allocate) (wait) (return) (measurement) (entanglement)}] {};
            \end{scope}

            \node (flavours) [below=of core-bg, yshift=5pt] {\textbf{NetQASM flavors:}};
            \path let
            \p1 = (wait.west), \p2=(entanglement.east)
            in node (vanilla) [flavour-box, below=of $(\p1)!1/8!(\p2)$]
                  {\textbf{Vanilla flavor:}\\init,\\x, y, z,\\h, s, k, t,\\rot\_x, rot\_y, rot\_z,\\cnot, cphase};
            \path let
            \p1 = (wait.west), \p2=(entanglement.east)
            in node (nv) [flavour-box, below=of $(\p1)!3/8!(\p2)$]
                  {\textbf{NV flavor:}\\init,\\rot\_x, rot\_y, rot\_z,\\cx\_dir, cy\_dir};
            \path let
            \p1 = (wait.west), \p2=(entanglement.east)
            in node (ti) [flavour-box, below=of $(\p1)!5/8!(\p2)$]
                  {\textbf{TI flavor:}\\\dots};
            \path let
            \p1 = (wait.west), \p2=(entanglement.east)
            in node (other-flavour) [flavour-box, below=of $(\p1)!7/8!(\p2)$, path fading=east]
                  {\textbf{\dots}};
      \end{tikzpicture}
      \caption{The \textbf{core} of \netqasm\ consists of eight groups of
            instructions. The quantum gates are defined as a set of software-visible
            gates part of a \netqasm\ \textbf{flavor}. The \textbf{vanilla flavor}
            is the unique platform-independent \netqasm\ \textbf{flavor} of
            \netqasm, which can be used by a compiler.}\label{fig:instructions}
\end{figure}
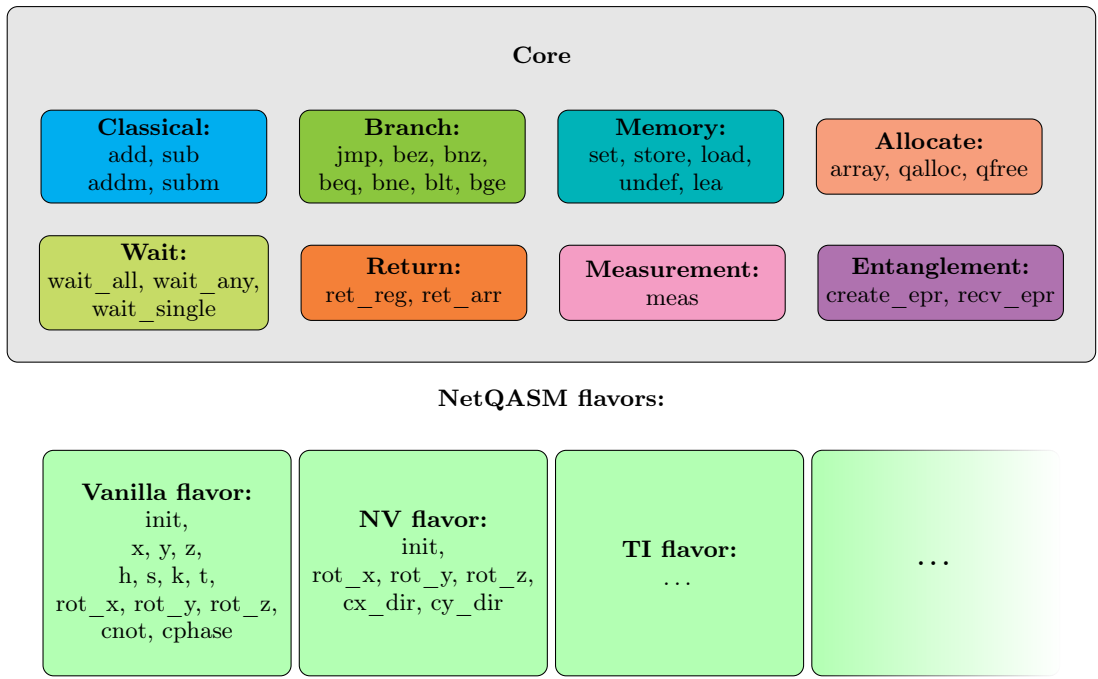

There are eight groups of
instructions in the \textbf{core} of \netqasm. Also summarized in
\cref{fig:instructions}, these are:
\begin{itemize}
      \item \textbf{Classical:} Classical arithmetic on integers.
      \item \textbf{Branch:} Branching operations for performing conditional
            logic.
      \item \textbf{Memory:} Read and write operations to classical memory (register and arrays).
      \item \textbf{Allocate:} Allocation of qubits and arrays.
      \item \textbf{Wait:} Waiting for certain events. This can for example be
            the event that entanglement has been generated by the network stack.
      \item \textbf{Return:} Returning classical values from the \QNPU to the
            \host. In our implementation we implement this by having the \QNPU
            write to the shared memory so that the \host can access it.
      \item \textbf{Measurement:} Measuring a qubit.
      \item \textbf{Entanglement:} Creating entanglement with a remote node using
            the quantum network stack.
\end{itemize}

Quantum gates are specific to a \netqasm\ \textbf{flavor} and given as a set of
software-visible gates of a given platform, see \cref{sec:design_considerations}. There is a
single platform-independent \netqasm\ \textbf{flavor} which we call the
\textbf{vanilla flavor}, see \cref{fig:instructions}. The \textbf{vanilla
      flavor} can be used as an intermediate representation for a compiler.

\subsection{Compilation}
Although application programmers could write \netqasm subroutines manually, and
let their (classical) application code send these subroutines to the \QNPU, it
is useful and more user-friendly to be able to write quantum internet
applications in a higher level language, and have the quantum parts compiled to
\netqasm subroutines automatically. For this, we use the compilation steps
depicted in \cref{fig:comp_chain}. The format and compilation of the
higher-level programming language is not part of the \netqasm specification.
However, we do provide an implementation in the form of an SDK, see
\cref{sec:python-sdk}.

\tikzstyle{box} = [rectangle, rounded corners, minimum width=6cm, minimum height=1cm, text centered, draw=black]
\tikzstyle{select} = [rectangle, dotted, draw=blue, very thick, inner sep=0.2cm]
\tikzstyle{arrow} = [thick,->,>=stealth]
\begin{figure*}
      \centering
      \begin{tikzpicture}[node distance=2cm]
            \node (higher) [box, fill=red] {Higher-level programming language};
            \node (hybrid) [box, below of=higher, fill=red] {Hybrid quantum-classical program};
            \node (netqasm-unop) [box, below of=hybrid, fill=black!10!orange] {NetQASM (vanilla flavor)};
            \node (netqasm-op1) [box, below of=netqasm-unop, fill=black!10!yellow] {NetQASM (HW flavor)};
            \node (netqasm-op2) [box, below of=netqasm-op1, fill=black!10!yellow, yshift=-1cm] {NetQASM (HW flavor)};
            \node[anchor=west,inner sep=3pt,text width=6cm] at ($(higher.east) + (15pt,0)$) {Full application, written by a programmer (e.g. Python)};
            \node[anchor=west,inner sep=3pt,text width=6cm] at ($(hybrid.east) + (15pt,0)$) {Full application, multiple subroutines compiled together, including also classical logic at application level};
            \node[anchor=west,inner sep=3pt,text width=6cm] at ($(netqasm-unop.east) + (15pt,0)$) {Virtual qubits part of allocated unit-module, abstract gates, single subroutine};
            \node[anchor=west,inner sep=3pt,text width=6cm] at ($(netqasm-op1.east) + (15pt,0)$) {Same as above except for software-visible gates (platform-dependent)};
            \node[anchor=west,inner sep=3pt,text width=6cm] at ($(netqasm-op2.east) + (15pt,0)$) {Input from above};
            \draw [arrow] (higher) -- node[anchor=east] {} (hybrid);
            \draw [arrow] (hybrid) -- node[anchor=east] {(multiple subroutines)} (netqasm-unop);
            \draw [arrow] (netqasm-unop) -- node[anchor=east] {(heavy compiling)} (netqasm-op1);
            \draw [arrow] (netqasm-op1) -- node[anchor=east] {} (netqasm-op2);
            \node (compilation) [select,fit={($(hybrid.west) + (-5pt,0)$) ($(hybrid.north) + (0,8pt)$) (netqasm-unop) (netqasm-op1)}] {};
            \node[anchor=north west,inner sep=3pt] at (compilation.north west) {Compilation};
            \node (applayer) [select,fit={($(higher.north) + (0,8pt)$) (compilation)}] {};
            \node[anchor=north west,inner sep=3pt] at (applayer.north west) {Application layer};
            \node (qnpu) [select,fit={($(netqasm-op2.north) + (0,8pt)$) (netqasm-op2)}] {};
            \node[anchor=north west,inner sep=3pt] at (qnpu.north west) {QNPU};
      \end{tikzpicture}
      \caption{Compilation steps from higher-level programming language, to the
            \netqasm\ \textbf{flavor} exposed by the specific platform. What is
            contained at each level is further specified to the right of the
            diagram.}
      \label{fig:comp_chain}
\end{figure*}
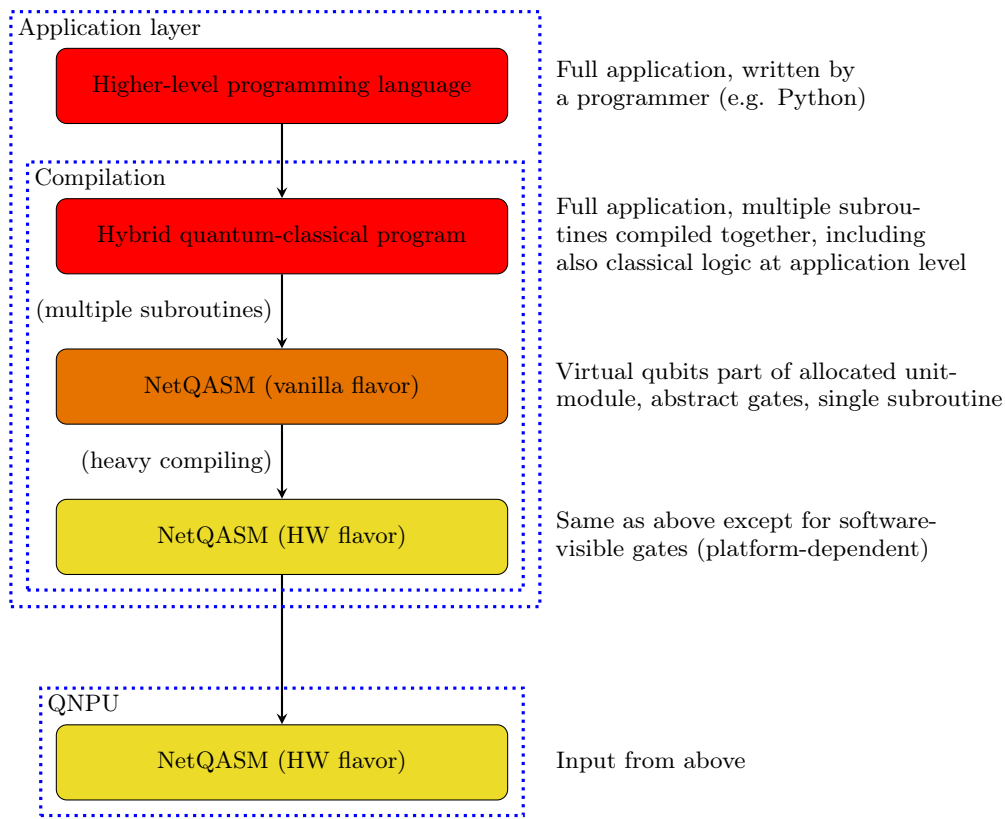

\section{Python SDK}\label{sec:python-sdk} We implemented \netqasm\ by
developing a Software Development Kit (SDK) in Python. This
SDK allows a programmer to write quantum network programs as Python code,
including the quantum parts. These parts are automatically translated to NetQASM
subroutines. The SDK contains a simulator that simulates a quantum network
containing end-nodes, each with a \QNPU. The SDK can execute programs by
executing their classical parts directly and executing the quantum parts as
\netqasm subroutines on the simulated \QNPU. By executing multiple programs at
the same time, on the same simulated network, a whole multi-partite application
can be simulated. In~\cref{sec:evaluation} we use this SDK to evaluate some of
the design decisions of \netqasm.

We refer to the docs at~\cite{git_netqasm} for the latest
version of the SDK. Below, we give an example of an application written in the
SDK to give an idea of how development in the SDK looks like. In
\cref{app:examples_sdk} we provide a few more examples of applications in the
SDK and their corresponding \netqasm\ subroutines.

All code can be found at~\cite{git_netqasm}
and~\cite{git_squidasm}, including: (1) Tools for serializing (de-serializing)
to (from) both human-readable text form and binary encoding,
(2) the \netqasm\ SDK, together with compilers (no optimization yet),
(3) support for running applications written in the SDK on the simulators
\netsquid~\cite{netsquid,coopmans2021netsquid} and
\simulaqron~\cite{dahlberg2018simulaqron}, and
(4) implemented applications in \netqasm, including: anonymous
transmission~\cite{Christandl2005anonymous},  BB84~\cite{bb84},  blind
quantum
computing~\cite{broadbent2009universal,fitzsimons2017unconditionally},  CHSH
game~\cite{Kaniewski2016},  performing a distributed
CNOT~\cite{denchev2008distributed},  magic square
game~\cite{brassard1999magicsquare},
teleportation~\cite{bennett1993teleporting}.

\subsection{SDK}\label{sec:sdk} The SDK of \netqasm\ uses a similar framework to
the SDK used by the predecessor \cqc~\cite{git_cqc}. Any program on a node
starts by setting up a \py{NetQASMConnection} to the \QNPU-implementation in the
\emph{backend}. The \py{NetQASMConnection} encapsulates all communication that
the \host\ does with the \QNPU. More information about supported backends can be
found below in \cref{sec:backends}. Using the \py{NetQASMConnection} one can for
example construct a \py{Qubit} object. The \py{Qubit} object has methods for
performing quantum gates and measurements. When these methods are called,
corresponding \netqasm\ instructions are included in the current subroutine
being constructed. One marks the end of a subroutine, and the start of another,
either by explicitly calling \py{flush} on the \py{NetQASMConnection} or by ending
the scope of the \nq{with NetQASMConnection ...} context.

The following Python code shows a basic application written in the \netqasm\
SDK. The application will be compiled into a single subroutine executed on the
\QNPU, which creates a qubit, performs a Hadamard operation, measures the qubit
and returns the result to the application layer.
\begin{pycode}
  # Setup connection to backend
  # as the node Alice
  with NetQASMConnection("Alice") as alice:
    # Create a qubit
    q = Qubit(alice)
    # Perform a Hadamard on the qubit
    q.H()
    # Measure the qubit
    m = q.measure()
    # The end of the context also marks
    # the end of the subroutine
    # automatically but can also be done
    # explicitly using `alice.flush()`
\end{pycode}

The following \netqasm\ subroutine is the result of translating the above
Python code to \netqasm of the vanilla (platform-independent) flavor.

\begin{nqcode}
  # NETQASM 1.0
  # APPID 0
  // Set the virtual qubit ID to use
  set Q0 0

  // Allocate and initialize a qubit
  qalloc Q0
  init Q0

  // Perform a Hadamard gate
  h Q0

  // Measure the qubit
  meas Q0 M0

  // Return the outcome
  ret_reg M0
\end{nqcode}

\subsubsection{Backends}\label{sec:backends} As mentioned above, the
\py{NetQASMConnection} in the SDK is responsible for communicating with the
implemented \QNPU\ in the \emph{backend}. The \emph{backend} can either be a
simulator or an actual \QNPU\ using real quantum hardware. Currently supported
backends are the simulators \squidasm~\cite{git_squidasm} (using
\netsquid~\cite{netsquid, coopmans2021netsquid}) and
\simulaqron~\cite{dahlberg2018simulaqron}. A physical implementation of \QNPU\
running on quantum hardware is being worked on at the time of writing. Using the
SDK provided at~\cite{git_netqasm}, one can for example simulate a set of
program files for the nodes of a quantum network on \netsquid\ using a density
matrix formalism with the command:
\begin{nqcode}
  netqasm simulate --simulator=netsquid --formalism=dm
\end{nqcode}
For more details see the docs at~\cite{git_netqasm}.

\newcommand{\figscale}{1.0}

\section{Evaluation}\label{sec:evaluation} We evaluate two of the design choices
that we made for \netqasm: (1) exposing unit-modules to the \host and (2) adding
the possibility to use platform-specific flavors of instructions. For both
elements we study the difference in including them in \netqasm versus not
including them. We do this by simulating a teleportation application and a blind
quantum computation application. These examples also showcase the ability of
\netqasm to express general quantum internet applications.

We have implemented a simulator, called \squidasm~\cite{git_squidasm}, that
simulates a network in which end-nodes have the internal architecture as
described in~\cref{sec:preliminaries}, that is, with an \host and a \QNPU. The
simulator internally uses NetSquid~\cite{netsquid}, which was made specifically for the
simulation of quantum networks. \squidasm executes programs written using the SDK
(\cref{sec:python-sdk}), including sending \netqasm subroutines to the
(simulated) \QNPU.

We evaluate the performance of \netqasm by looking at the runtime quality of two
applications, both consisting of two programs (one per node). The first is a
teleportation of a single qubit from a sender node to a receiver node. We define
the quality as the fidelity between the original qubit state at the sender and
the final qubit state at the receiver. The second application is a blind
computation protocol which involves a client and a server. The server
effectively performs, blindly, a single-qubit computation on behalf of the
client. The protocol is a so-called \textit{verifiable blind quantum
  computation}~\cite{fitzsimons2017unconditionally}. This means that some of the
rounds of the protocols are \textit{trap rounds}. We define the quality that we
evaluate as the error rate of these trap rounds, since this indicates
the blindness of the server.

We run these applications on \squidasm, where we simulate realistic quantum
hardware. Specifically, we simulate nodes based on nitrogen-vacancies (NV) in
diamond, that can do heralded entanglement generation between each other. The
simulated hardware uses noise models that are also used
in~\cite{coopmans2021netsquid}. For more details, see~\cref{app:simulation}.

\subsection{Unit modules}\label{sec:evaluation-unit-modules}
We ask ourselves the question whether it pays off to expose unit modules, that
is, a qubit topology with gate- and entanglement information. Specifically, we
want to know if there are situations where knowing the unit module gives the
\host an opportunity to optimize the application in a way that is not possible
when not knowing the unit module. If so, we are interested in how much advantage
this gives (in terms of the runtime quality defined above).

In the next section we show that there are indeed situations where knowledge of
the unit module is advantageous. It can be that the order in which \netqasm
instructions are issued in a subroutine is sub-optimal, since virtual qubit IDs
may be mapped in such a way that the \QNPU has to move virtual qubits to
different physical qubits in order to execute the instructions. If the \host
layer does not know this mapping, it cannot know that the instructions are
ordered sub-optimally. With knowledge of the unit module, on the other hand,
the \host can optimize the order and the overall application performance is
improved.

We consider a teleportation application where a \textit{sender} program
teleports a single qubit to another \textit{receiver} program. It is assumed
that the underlying platform is based on nitrogen-vacancy centers in diamond
(NV) and use well-established models for both the noise and operations supported
on such platforms, see \cref{app:simulation}. The sender program uses two
qubits: one to create entanglement with the receiver (qubit E), and one to send
(teleport) to the receiver (qubit T). At some point, the sender measures both
qubits, after which it sends the outcomes to the receiver so that it can do the
relevant corrections on its received qubit. We assume that the sender
program is written in a higher-level language like, like in our SDK
(\cref{sec:sdk}), and in such a way that it first issues a measurement operation
on qubit T, and then on E. However, due to the differences in characteristics of
the physical qubits, as will be explained below, it is more efficient to first
do the measurement on E, and then on T. Now we consider two scenarios, namely
\begin{itemize}
  \item \textbf{Unit-modules (UM)}. We assume that the sender program is
        written and executed on a software stack implementating \netqasm,
        which means that the application's view of its quantum working memory
        is in the form of a unit module. This unit module contains information
        about the above-mentioned hardware restrictions, and therefore a
        compiler can take advantage of it by re-ordering the measurement
        operations while generating the \netqasm subroutines to be sent to the
        \QNPU.
  \item \textbf{No unit-modules (NUM)}. In this case the software stack also
        implements \netqasm, but without unit modules. Specifically, the
        application sees its quantum memory as just a number of uniform
        qubits. Therefore, a compiler for this application does not know about
        the hardware restrictions, and will construct \netqasm-subroutines
        sent to the \QNPU\ without doing any optimization and leaves the order
        of the operations to be performed as they are specified in the
        high-level SDK.
\end{itemize}

Let's first go through the steps of the teleportation application:\newline
\begin{description}
  \item[\textit{sender}]:
        \begin{enumerate}
          \item Initialize qubit $q_t$ to be teleported in a Pauli state.
          \item Create entanglement with \textit{receiver} using qubit
                $q_s$.
          \item Perform CNOT gate with $q_t$ as control and $q_s$ as target.
          \item Perform Hadamard gate on $q_t$.
          \item Measure qubit $q_t$ and store outcome as $m_1$.
          \item Measure qubit $q_s$ and store outcome as $m_2$.
          \item Send $m_1$ and $m_2$ to \textit{receiver}.
        \end{enumerate}
  \item[\textit{receiver}]:
        \begin{enumerate}
          \item Receive entanglement with \textit{sender} using qubit $q_r$.
          \item Receive measurement outcomes from \textit{sender}.
          \item Apply correction operations on $q_r$ based on measurement
                outcomes.
        \end{enumerate}
\end{description}

We will now consider the order of the steps of the \textit{sender}. Firstly, we
assume that the qubit to be teleported, $q_t$, is always created before the
entanglement. We motivate this assumption below. For this reason, steps 1--3 and
7 are fixed and cannot change. However, we are free to do step 6 before step 4
and 5, since these single-qubit operations and measurements commute, as long as
we are consistent with the outcomes $m_1$ and $m_2$. Let's now consider what
impact this decision of measuring $q_s$ before $q_t$ or not has on the quality
of execution for a NV-platform.

One of the biggest restrictions on a NV-platform is the topology of the qubits.
In particular, the NV-platform has a single communication-qubit (electron)
surrounded by some number of storage qubits (carbon spins), see for example
\cref{fig:topology}. The single communication qubit is not only responsible for
any remote entanglement generation but also for any two-qubit gate and is the
only qubit that can be directly measured. These restrictions require qubit
states to be moved back and forth between the communication qubit and the
storage qubits in order to free up the communication qubit, to create new
entanglement or to measure another qubit. Since the operation of moving a qubit
state is relatively slow on this platform (up to a
millisecond~\cite{Humphreys2018}) and adds noise to the qubits, it is important
to try to minimize the number of moves needed. For more details on the
NV-platform, see for example~\cite{Bernien2014} or~\cite{dahlberg2019linklayer}.

In the steps of the \textit{sender} above, the communication qubit is first
initialized to a Pauli state. This state is then moved to a storage qubit to
free up the communication qubit in order to create entanglement with the
\textit{receiver}. Then in step 5, $q_t$ should be measured, which is currently
in the storage qubit. This requires the qubit state to first be moved to the
communication qubit. However, at this point the communication qubit is occupied
by the entangled pair and therefore first needs to be moved to a second storage
qubit. Qubit $q_t$ can then be moved to the communication qubit to be measured
and then the same is done for $q_s$, requiring in total four move operations and
three physical qubits.

We can now see that performing step 6 before 4 and 5 has the advantage that this
qubit is already in the communication qubit and can be measured directly without
moving it first. Afterwards, $q_t$ can be moved to the communication qubit,
which is cleared after the measurement, requiring in total only 2 move
operations and only two physical qubits. The decision of performing step 6
before 4 and 5 is highly dependent on the NV-platform and can only be made by a
compiler that is aware about these restrictions. The inclusion of unit-modules
and qubit types in the \netqasm-framework, which are exposed to the compiler at
the \host, allows for these optimization decision and can therefore improve the
quality of execution.

\begin{figure}
  \centering
  \subcaptionbox{\label{fig:sweep_gate_noise}}{%
    \includegraphics[scale=0.8]{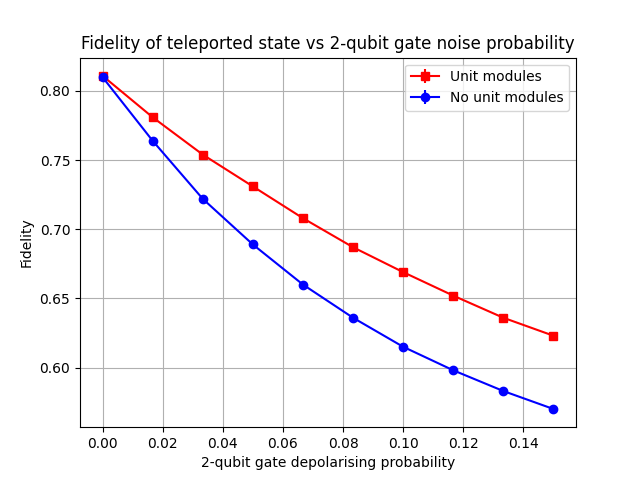}}
  \subcaptionbox{\label{fig:sweep_gate_time}}{%
    \includegraphics[scale=0.8]{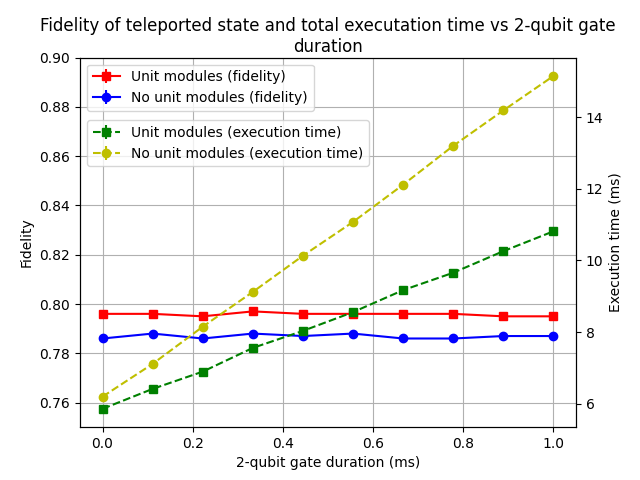}}

  \caption{(a) Average fidelity between the original state at the sender and
    the final state at the server, as a function of the depolarizing noise
    of the native two-qubit gate of the NV-platform, both for the case of
    performing step 6 after (\textbf{No unit modules}) and before
    (\textbf{Unit modules}) step 4 and 5. Execution time of the native
    two-qubit gate is set to 0.5 ms. The rest of the parameters used are
    listed in \cref{app:simulation}. Each point is the average over each of
    the six Pauli states as initial state, repeated 100 times. (b) Average
    fidelity of the teleported state (left y-axis, solid lines) and total
    execution time of the teleportation application (right y-axis, dashed
    lines) as a function of the execution time of the native two-qubit gate
    of the NV-platform, both for the case of performing step 6 after
    (\textbf{No unit modules}) and before (\textbf{Unit modules}) step 4 and
    5. Dephasing parameter of the native two-qubit gate is set to 0.02. The
    rest of the parameters used are listed in \cref{app:simulation}. Each
    point is the average over each of the six Pauli state as initial state,
    repeated 100 times. In both figures, error bars are smaller than the drawn
    dots.}

\end{figure}

For the two scenarios we consider, i.e. performing step 6 before 4 and 5
(\textbf{Unit modules (UM)}) or not (\textbf{No unit-modules (NUM)}), we check
the average fidelity of the teleported state as a function of the gate noise
(\cref{fig:sweep_gate_noise}), as well as the average fidelity and execution
time as a function of gate duration (\cref{fig:sweep_gate_noise}), of the native
two-qubit gate of the NV-platform. We see that performing step 6 before 4 and 5
improves both total execution time and average fidelity. This can be explained
by the fact that using unit modules allowed a compiler to produce \netqasm code
containing fewer two-qubit gates. Therefore, an increase in two-qubit gate noice
leads to a lower fidelity. Also, an increase in two-qubit gate duration leads to
higher execution time difference between the two scenarios. Finally,
\cref{fig:sweep_gate_noise} shows that the two-qubit gate duration does not
affect the final fidelity in this situation, but the difference between using
unit modules versus not using them remains.

\subsection{Flavors}\label{sec:evaluation-flavours}
While aiming to let \netqasm be mostly platform-independent, we did also choose
to allow platform-specific instructions, bundled in flavors. The idea is that
this allows for platform-specific optimization leading to better application
performance. Here we evaluate if flavors really impact potential performance,
and if so how much.

We show that platform-specific optimization can indeed improve
application performance, and that there are such optimizations that are not
possible without flavors. We see that it has impact mostly on the execution time,
but not necessarily on outcome quality.

\begin{figure}
  \centering
  \includegraphics[width=\textwidth]{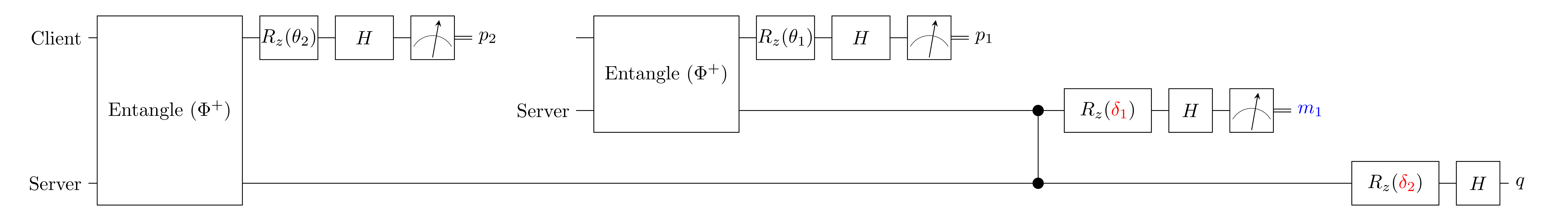}
  \caption{Circuit representation of the simulated BQC application. The client
    remotely prepares two qubits on the server, by twice creating an
    entangled pair with the server followed by a local measurement. The
    server locally entangles its two qubits (cphase gate). Then, the client
    and server use classical communication to further guide the server's
    quantum operations. The client computes $\delta_1 = \alpha - \theta_1 +
      p_1 \cdot \pi$ and sends this to the server. The server uses the
    received value to do a local rotation and later sends measurement
    outcome $m_1$ back to the client. The client then sends $\delta_2 =
      (-1)^{m_1} \cdot (\beta - \theta_2 + p_2 \cdot \pi)$ to the server.
    The qubit state $q$ is the result of this application.
  }
  \label{fig:bqc_app}
\end{figure}

We consider the blind computation application depicted in \cref{fig:bqc_app}, where both
the client and server node implement the NV hardware. Again we compare two scenearios,
in this case:
\begin{itemize}
  \item \textbf{Vanilla}. We compile both the client's and server's
        application code to \netqasm subroutines with the vanilla flavor. The
        \QNPU, controlling NV hardware which does not implement all vanilla
        gates natively, needs to translate the vanilla instructions on the go.
        We assume this translation is ad-hoc and does not do any optimizations
        like removing redundant gates.
  \item \textbf{NV}. The code is compiled to \netqasm subroutines containing
        instructions in the NV flavor, and redundant gates are optimized
        away. The \QNPU can directly execute the instructions on the hardware.
\end{itemize}

We implemented this by writing two separate programs in the SDK, one for the
client and one for the server. The SDK automatically compiles the relevant parts
of these programs into \netqasm subroutines. Classical communication (values
$\delta_1$, $m_1$ and $\delta_2)$ is done purely between the two simulated
application layers, so these operations are not compiled to \netqasm subroutines.
More details about the simulation can be found in \cref{app:simulation}.

The protocol is a verifiable blind quantum protocol~\cite{fitzsimons2017unconditionally},
which means that the circuit in~\cref{fig:bqc_app} is run multiple times, namely once per round.
Some of these rounds are \textit{trap rounds} in which the client chooses a special set of input values.
Such a trap round can either succeed or fail, depending on the values returned by the server.
The fraction of trap rounds that fail is called the error rate.
The error rate should stay low in order for the computation to be blind.

\begin{figure}
  \centering
  \includegraphics[scale=0.8]{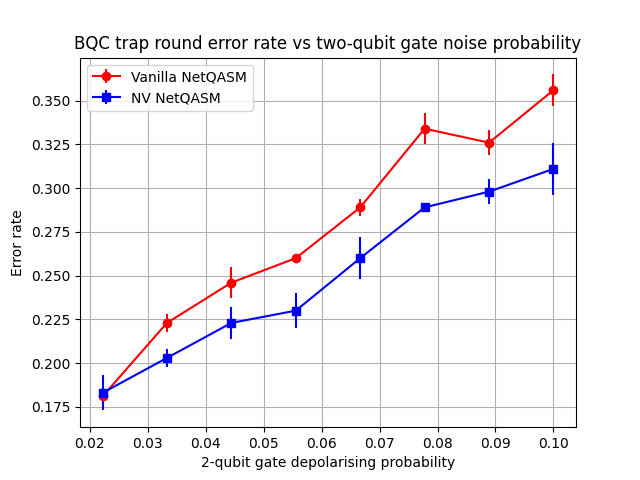}
  \caption{
    Average error rate of trap rounds for the circuit of~\cref{fig:bqc_app}.
    Each point is the average over four combinations of $\theta_1$ and $\theta_2$,
    each used in 500 trap rounds. It can be seen that using the vanilla (platform-independent)
    \netqasm flavor results in a worse (higher) error rate on average.}
  \label{fig:plot_bqc}
\end{figure}

We simulate the BQC application by running the client's and server's programs in \squidasm.
We look at the error rate of the trap round as a function of the two-qubit gate noise.
The result can be seen in~\cref{fig:plot_bqc}. It can be seen that
using the NV flavor provides a better (lower) error rate than using the vanilla flavor.
This can be explained by noting that \netqasm instructions in the vanilla flavor
are mapped ad-hoc to native NV gates by the \QNPU at runtime, which leads to
more two-qubit gates in total.

\subsection{Relation to other results}
We note that a similar question of how many physical details to expose from
lower-level layers (in our case the \QNPU) to higher-level layers (in our case
the \host) has also been evaluated in~\cite{murali2019fullstack}. Their
conclusion is that exposing and leveraging some of these details can indeed
improve certain program success metrics. That result agrees with that of ours,
which shows that program execution quality can improve by exposing and leveraging
unit modules and platform-specific \netqasm flavors.
\section{Conclusion}\label{sec:conclusion} \netqasm enables the development of
quantum internet applications in a platform-independent manner. It solves the
question of dealing with the complexity of having both classical and quantum
operations in a single program, while at the same time providing a relatively
simple format for \QNPU-like layers to handle. Multiple applications, such as
remote teleportation and blind quantum computation, have already been
implemented. A simple compiler has been implemented that can translate code
written in the higher-level SDK into \netqasm.

Additionally to the work in this paper, we are also developing a physical
implementation of the \QNPU. One key component in this implementation is the
\emph{Quantum Node Operating System} (\qnodeos), which acts as the bridge
between the applications and the physical layer. \qnodeos\ will be presented in
a dedicated paper including results of a first integration test between
\netqasm, \qnodeos\ and underlying physical quantum hardware. This will mark the
first time a quantum network node has been programmed using platform-independent
code.

\section*{Acknowledgments}

We thank Arjen Rouvoet and \"Onder Karpat for valuable discussions. This work was supported by ERC
Starting Grant, EU Flagship on Quantum Technologies, Quantum Internet Alliance,
NWO VIDI.

\bibliographystyle{unsrt}
\bibliography{refs}

\appendix

\section{Flow of messages}\label{sec:app-messages}
Here we define the content of each of the messages being sent between the \host
and the \QNPU. Each message has an ID chosen by the \host\ which is used to
associate replies from the \QNPU\ to the \host.
\begin{itemize}
  \item \nq{RegisterApp}: Sent once from the \host\ to the \QNPU\ whenever a
        new application starts. Contains information on what resources are
        required by the application, in particular:
        \begin{itemize}
          \item \nq{unit_module_spec}: Specification of unit-module needed,
                e.g. number of qubits.
          \item \nq{epr_socket_spec}: Specification of EPR sockets needed,
                see~\cite{kozlowski2020networklayer}, containing (1) EPR socket ID,
                (2) remote node ID, (3) remote EPR socket ID and (4) minimum
                required fidelity.
        \end{itemize}
  \item \nq{RegisterAppOK}: Returned from the \QNPU\ when application is
        registered, containing an application ID to be used for future messages.
        \begin{itemize}
          \item \nq{app_id}: Application ID.
        \end{itemize}
  \item \nq{RegisterAppErr}: Returned from the \QNPU\ when registration of
        application failed. For example if required resources could not be met.
        \begin{itemize}
          \item \nq{error_code}: Error code specifying what went wrong.
        \end{itemize}
  \item \nq{Subroutine}: Message from the \host\ to the \QNPU, containing a
        subroutine to be executed. Details on the content are presented in later
        sections.
        \begin{itemize}
          \item \nq{app_id}: Application ID.
          \item \nq{subroutine}: The subroutine to be executed.
        \end{itemize}
  \item \nq{Done}: Message from the \QNPU\ to the \host, indicating that a
        subroutine has finished. Which subroutine is indicated by the message
        ID.
        \begin{itemize}
          \item \nq{message_id} Message ID used for the
                \nq{Subroutine}-message.
        \end{itemize}
  \item \textbf{Update memory}: The \host\ will have access to a copy of the
        memory allocated by the \QNPU\ for certain registers and arrays, see
        \cref{sec:language}. This memory is read-only by the \host. Updates to the
        copy of the memory are performed by the end of a subroutine or if the
        subroutine is waiting. Furthermore, updates need to be explicitly specified
        in the subroutine by using one of the return-commands. How the actual update
        is implemented depends on the platform and can either be done by
        message-passing or with an actual shared memory. However, the subroutine is
        independent from this implementation. The \host\ will be notified by an
        explicit message whenever the memory is updated.
  \item \nq{StopApp}: Sent from the \host\ to the \QNPU\ indicating that an
        application is finished.
\end{itemize}

\section{Operands}\label{sec:operands}
In this section we give the exact definition of the types of operands used in the \netqasm\ language.
Each instruction of \netqasm\ takes one or more operands.
There are five types of operands, which are listed and described below.
Each instruction has a fixed types of operands at each position.
The exact operands for each instruction is listed in \cref{app:instructions}.
We note also that in the human-readable text-form of \netqasm, there are also \textit{branch variables}.
However, these are always replaced by \IMMEDIATE{}s (constants), corresponding to the instruction number of the subroutine, before serializing, see \cref{sec:branch_variables}.

The operand types of \netqasm\ are:
\begin{itemize}
  \item \IMMEDIATE (constant): An integer seen as it's value.
        The following instruction, \nq{beq} \emph{branch-if-equal}, branches to instruction index \nq{12} since the number \nq{0} equals the number \nq{0}.
        \begin{nqcode}
beq 0 0 12\end{nqcode}
        In the binary encoding used at~\cite{git_netqasm}, \IMMEDIATE{}s are \nq{int32}.
  \item \REGISTER: A register specifying a register name and a index.
        The following instruction sets index \nq{0} of the register name \nq{R} to be \nq{0}.
        \begin{nqcode}
set R0 0\end{nqcode}
        In the current version of \netqasm\ there are four register names and the indices are relative to the names.
        They are all functionally the same but are meant to be used for different purposes and increase readability:
        \begin{itemize}
          \item \nq{C}: Constants, meant to only be \nq{set} once throughout a subroutine.
          \item \nq{R}: Normal register, used for looping etc.
          \item \nq{Q}: Stores virtual qubit IDs.
          \item \nq{M}: Stores measurement outcomes.
        \end{itemize}
        In the binary encoding used at~\cite{git_netqasm}, \REGISTER{}s are specified by one byte and hold one \nq{int32}.
  \item \ADDRESS: Specifies an address to an array.
        Starts with \nq{@}.
        The following instruction declares an array of length \nq{10} at address \nq{0}.
        \begin{nqcode}
array 10 @0\end{nqcode}
        For more information about arrays, see below.
        The address here is just an identifier of the array and does not refer to a actual memory address.
        For this reason \nq{@1} above does not mean the second entry of the declared array but simply a different array.
        Addresses are relative to the application ID and are valid across subroutines.
  \item \textbf{ARRAY\_ENTRY}: Specifies an entry in an array.
        Takes the form \nq{@a[i]}, where \nq{a} specifies the address and \nq{i} the index.
        The following instruction stores the value of \nq{R0} to the second entry of the array with address \nq{0}.
        \begin{nqcode}
store R0 @0[1]\end{nqcode}
        In the text-form \nq{i} can either be an \IMMEDIATE\ or a \REGISTER, however in the binary encoding used at~\cite{git_netqasm}, \nq{i} is always a \REGISTER.
        This is handled by the compiler by using a \nq{set}-command before.
  \item \textbf{ARRAY\_SLICE}: Specifies a slice of an array.
        Takes the form \nq{@a[s:e]}, where \nq{a} specifies the address, \nq{s} the start-index (inclusive) and \nq{e} the end-index (exclusive).
        The following instruction waits for the second to the fourth entry of array with address \nq{0} to become not \nq{null}, see \cref{app:waiting}.
        \begin{nqcode}
wait_all @0[1:4]\end{nqcode}
        In the text-form \nq{s} and \nq{e} can either be an \IMMEDIATE{}s or a \REGISTER{}s, however in the binary encoding defined used at~\cite{git_netqasm}, \nq{s} and \nq{e} are always a \REGISTER{}s.
        This is handled by the compiler by using a \nq{set}-commands before.
\end{itemize}

\section{Branch variables}\label{sec:branch_variables}
The human-readable text-form of \netqasm\ supports the use of \emph{branch variables}.
Branch labels are declared as \nq{VAR}: before the instruction to branch to.
Before serializing a \netqasm-subroutine, all branch variables are replaced with \IMMEDIATE{}s corresponding to the correct \emph{instruction index}.
Delaying this replacement to the end is useful if the compiler wants to move around instructions.
For example if a subroutine is as follows:
\begin{nqcode}
# NETQASM 1.0
# APPID 0
set R0 0

// Loop entry
LOOP:
beq R0 10 LOOP_EXIT

// Loop body
// If statement
bge R0 5 ELSE
// true block
add R0 R0 1
jmp IF_EXIT
// false block
ELSE:
add R0 R0 2
IF_EXIT:

// Loop exit
jmp LOOP
LOOP_EXIT:\end{nqcode}
Which effectively does the same as the following program written in Python (where the variable \py{i} corresponds to the register \nq{R0} above).
\begin{pycode}
i = 0
while i != 10:
  if i < 5:
    i += 1
  else:
    i += 2
\end{pycode}
After replacing the branch labels the body of the subroutine will instead look:
\begin{nqcode}
store R0 0
beq R0 10 7
bge R0 5 5
add R0 R0 1
jmp 6
add R0 R0 2
jmp 1\end{nqcode}

\section{Arrays}
Classical data produced during the execution of a subroutine are stored in either fixed registers or allocated arrays.
Arrays in \netqasm\ have fixed-length, which is specified when declared using the \nq{array}-instruction.
Each entry of an array is an \emph{optional} \IMMEDIATE, meaning that the entry is an integer (e.g. \nq{int32}) or not defined (\nq{null}).
The arrays can be used to collect measurement outcomes to be returned to the \host\ but also other data such as information about the generated remote entanglement~\cite{dahlberg2019linklayer,kozlowski2020networklayer}.
All wait-instructions of \netqasm\ wait for one or more entries in an array to become defined (i.e. not \nq{null}).
The main use-case is for the execution of the subroutine to wait until the quantum network stack of the \QNPU\ has finished generated an entangled pair with a remote node.
The subroutine will be waiting for information about the entangled pair to be stored in a given array.
Once this is done, the execution can proceed.

The following subroutine for example creates and array with three elements, stores the values \nq{1} and \nq{2} to the array and reads them and adds them up, storing the value in the third entry.
\begin{nqcode}
// Create two constant registers
set C1 1
set C2 2
// Make an array of three entries
array 3 @0
// Load the constants to the array
store C1 @0[0]
store C2 @0[1]
// Load the array entries to two other registers
load R0 @0[0]
load R1 @0[1]
// Add the registers and store the result in the first
add R0 R1 R0
// Store the sum in the third entry of the array
store R0 @0[2]\end{nqcode}

\section{Qubit address operands}
Commands that perform actions on qubits have \REGISTER-operands which specify the virtual address of the qubit to act on.
It is good practice to use register name \nq{Q} for these registers.
The following subroutine performs a Hadamard gates on qubits with virtual addresses \nq{0}, \nq{1} and \nq{2}.
\begin{nqcode}
set Q0 0
set Q1 1
h Q0
h Q1
set Q0 2
h Q0\end{nqcode}
Note that \nq{Q0} is used twice but the value of the register is different.

\section{Instructions}\label{app:instructions}
\setlist[itemize]{leftmargin=*}
Here we list the current instructions part of the \textbf{vanilla flavor} of \netqasm.
For the most up to date version of the language, refer to~\cite{git_netqasm}.
Commands are specified as follows:
\begin{itemize}
  \item \nq{name}: Description of instruction.
        \begin{enumerate}
          \item \IMMEDIATE : Description of op1
          \item \REGISTER : Description of op2
        \end{enumerate}
\end{itemize}

where \nq{name} is the name of the instruction, followed by the list of operands, specified by their type and description.
We note that in the human-readable text-form of \netqasm, it is allowed to provide an \IMMEDIATE\ for operands that are specified as \REGISTER.
The compiler will then replace these, using the \nq{set}-command.

\subsection{Allocation}

\begin{itemize}
  \item \nq{qalloc}: Start using a qubit in the unit module.
        \begin{enumerate}
          \item \REGISTER: The virtual address of the qubit.
        \end{enumerate}
  \item \nq{array}: Creates an array of a certain length (width is fixed)
        \begin{enumerate}
          \item \IMMEDIATE: Number of entries in the array.
          \item \ADDRESS: Address of array
        \end{enumerate}
\end{itemize}

\subsection{Initialization}

\begin{itemize}
  \item \nq{init}: Initializes a qubit to $|0\rangle$
        \begin{enumerate}
          \item \REGISTER: The virtual address of the qubit.
        \end{enumerate}
  \item \nq{set}: Set a register to a certain value.
        \begin{enumerate}
          \item \REGISTER: The register to assign a value to.
          \item \IMMEDIATE: The value to assign.
        \end{enumerate}
\end{itemize}

\subsection{Memory operations}
\begin{itemize}
  \item \nq{store}: Stores the value in a register to an index of an array.
        \begin{enumerate}
          \item \REGISTER: The register holding the value to store.
          \item \textbf{ARRAY\_ENTRY}: The array-entry to store the value to.
        \end{enumerate}
  \item \nq{load}: Loads the value from an index of an array to a register.
        \begin{enumerate}
          \item \REGISTER: The register to store the value to.
          \item \textbf{ARRAY\_ENTRY}: The array-entry holding the value.
        \end{enumerate}
  \item \nq{undef}: Sets an entry of an array to \nq{null}, see \cref{app:waiting}.
        \begin{enumerate}
          \item \textbf{ARRAY\_ENTRY} Array-entry to make \nq{null}.
        \end{enumerate}
  \item \nq{lea}: Loads a given address of an array to a register.
        \begin{enumerate}
          \item \REGISTER: The register to store the address to.
          \item \ADDRESS: The address to the array.
        \end{enumerate}
\end{itemize}

\subsection{Classical logic}

There are three groups of branch instructions: nullary, unary and binary.

\textbf{Nullary branching}
\begin{itemize}
  \item \nq{jmp}: Jump to a given line (unconditionally).
        \begin{enumerate}
          \item \IMMEDIATE: Line to branch to.
        \end{enumerate}
\end{itemize}

\textbf{Unary branching}
There are two unary branching instructions: \nq{beq} and \nq{bnz}, which both have the following structure:
\begin{itemize}
  \item \nq{b\{ez,nz\}}: Branch to a given line if condition fulfilled, see below.
        \begin{enumerate}
          \item \REGISTER: Value $v$ in condition expression.
          \item \IMMEDIATE: Line to branch to.
        \end{enumerate}
\end{itemize}

Branching occurs if:
\begin{itemize}
  \item \nq{bez}: $v = 0$ (branch-if-zero)
  \item \nq{bnz}: $v \neq 0$ (branch-if-not-zero)
\end{itemize}

\textbf{Binary branching}
There are four binary branch instructions: \nq{beq}, \nq{bne}, \nq{blt} and \nq{bge}, which all have the following structure:
\begin{itemize}
  \item \nq{b\{eq,ne,lt,ge\}}: Branch if condition fulfilled, see below.
        \begin{enumerate}
          \item \REGISTER: Value 1 $v_1$ in conditional expression.
          \item \REGISTER: Value 1 $v_1$ in conditional expression.
          \item \IMMEDIATE: Line to branch to.
        \end{enumerate}
\end{itemize}

Branching occurs if:
\begin{itemize}
  \item \nq{beq}: $v_0 = v_1$ (branch-if-equal)
  \item \nq{bne}: $v_0 \neq v_1$ (branch-if-not-equal)
  \item \nq{blt}: $v_0 < v_1$ (branch-if-less-than)
  \item \nq{bge}: $v_0 \geq v_1$ (branch-if-greater-or-equal)
\end{itemize}

\subsection{Classical operations}
There are currently four binary classical operations: addition (\nq{add}), subtraction (\nq{sub}) and addition (\nq{addm}), subtraction (\nq{subm}) modulo a number.
The first two have the following structure:
\begin{itemize}
  \item \nq{\{add,sub\}}: Perform a binary operation and store the result.
        \begin{enumerate}
          \item \REGISTER Register to write result ($r$) to.
          \item \REGISTER First operand in binary operation ($v_0$).
          \item \REGISTER Second operand in binary operation ($v_1$).
        \end{enumerate}
\end{itemize}

The second two have an additional operand to specify what module should be taken for the result:
\begin{itemize}
  \item \nq{\{add,sub\}m}: Perform a binary operation modulo \nq{mod} and store the result.
        \begin{enumerate}
          \item \REGISTER Register to write result ($r$) to.
          \item \REGISTER First operand in binary operation ($v_0$).
          \item \REGISTER Second operand in binary operation ($v_1$).
          \item \REGISTER Modulo in binary operation ($m$).
        \end{enumerate}
\end{itemize}

Binary operations are the following:
\begin{itemize}
  \item \nq{add}, $r = (v_0 + v_1)$
  \item \nq{sub}, $r = (v_0 - v_1)$
  \item \nq{addm}, $r = (v_0 + v_1) \pmod{m} $
  \item \nq{subm}, $r = (v_0 - v_1) \pmod{m} $
\end{itemize}

\subsection{Quantum gates}
\textbf{Single-qubit gates}
There is a number of single-qubit gates which all have the following structure
\begin{itemize}
  \item \nq{instr}: Perform a single-qubit gate.
        \begin{enumerate}
          \item \REGISTER: The virtual address of the qubit.
        \end{enumerate}
\end{itemize}

Single-qubit gates without additional arguments are the following.
\begin{itemize}
  \item \nq{x}: X-gate.
        \begin{equation}\label{eq:pauli_x}
          X = \begin{pmatrix} 0 & 1 \\ 1 & 0 \end{pmatrix}
        \end{equation}
  \item \nq{y}: Y-gate.
        \begin{equation}\label{eq:pauli_y}
          Y = \begin{pmatrix} 0 & -i \\ i & 0 \end{pmatrix}
        \end{equation}
  \item \nq{z}: Z-gate.
        \begin{equation}\label{eq:pauli_z}
          Z = \begin{pmatrix} 1 & 0 \\ 0 & -1 \end{pmatrix}
        \end{equation}
  \item \nq{h}: Hadamard gate.
        \begin{equation}
          H = \frac{1}{\sqrt{2}} \begin{pmatrix} 1 & 1 \\ 1 & -1 \end{pmatrix}
        \end{equation}
  \item \nq{s}: S-gate (phase)
        \begin{equation}
          S = \begin{pmatrix} 1 & 0 \\ 0 & i \end{pmatrix}
        \end{equation}
  \item \nq{k}: K-gate.
        \begin{equation}
          K = \frac{1}{\sqrt{2}} \begin{pmatrix} 1 & -i \\ i & -1 \end{pmatrix}
        \end{equation}
  \item \nq{t}: T-gate.
        \begin{equation}
          T = \begin{pmatrix} 1 & 0 \\ 0 & e^{i\pi/4} \end{pmatrix}
        \end{equation}
\end{itemize}

\textbf{Single-qubit rotations}
Additionally one can perform single-qubit rotations with a given angle.
The angles \nq{a} are specified by two integers \nq{n} and \nq{d} as:
\begin{equation}
  a = \frac{n\pi}{2^d}
\end{equation}
These instructions have the following structure
\begin{itemize}
  \item \nq{rot_\{x,y,z\}}: Perform a single-qubit rotation.
        \begin{enumerate}
          \item \REGISTER: The virtual address of the qubit.
          \item \IMMEDIATE: \nq{n}, for angle, see above.
          \item \IMMEDIATE: \nq{d}, for angle, see above.
        \end{enumerate}
\end{itemize}

Single-qubit rotations are the following.
\begin{itemize}
  \item \nq{rot_x}: Rotation around X-axis.
  \item \nq{rot_y}: Rotation around Y-axis.
  \item \nq{rot_z}: Rotation around Z-axis.
\end{itemize}

\textbf{Two-qubit gates}
There are two two-qubit gates which have the following structure
\begin{itemize}
  \item \nq{\{cnot,cphase\}}: Perform a two-qubit operation.
        \begin{enumerate}
          \item \REGISTER: The virtual address of the control qubit.
          \item \REGISTER: The virtual address of the target qubit.
        \end{enumerate}
\end{itemize}

Two-qubit gates are the following.
\begin{itemize}
  \item \nq{cnot}: Controlled $X$ gate.
  \item \nq{cphase}: Controlled $Z$ gate.
\end{itemize}

\textbf{Measurement}
\begin{itemize}
  \item \nq{meas}: Measure a qubit in the standard basis.
        \begin{enumerate}
          \item \REGISTER: The virtual address of the qubit.
          \item \REGISTER: The register to write outcome address to.
        \end{enumerate}
\end{itemize}

\textbf{Pre-measurement rotations}
To measure in other bases one can perform gates/rotations before the measurement.
If the same measurement basis is used a lot, one can also make use of pre-measurement rotations which can reduce the amount of communication needed internally in the \QNPU.
A pre-measurement rotations is specified by either the \nq{pmr_xyx}, \nq{pmr_zxz} or \nq{pmr_yzy} which have the following structure.
With any two of the bases X, Y and Z, one can do any rotation.
\begin{itemize}
  \item \nq{pmr_\{xyx,zxz,yzy\}}: Specify a pre-measurement rotation.
        \begin{enumerate}
          \item \IMMEDIATE: \nq{n0}, for angle of first rotation, see below.
          \item \IMMEDIATE: \nq{d0}, for angle of first rotation, see below.
          \item \IMMEDIATE: \nq{n1}, for angle of second rotation, see below.
          \item \IMMEDIATE: \nq{d1}, for angle of second rotation, see below.
          \item \IMMEDIATE: \nq{n2}, for angle of third rotation, see below.
          \item \IMMEDIATE: \nq{d2}, for angle of third rotation, see below.
        \end{enumerate}
\end{itemize}
If a pre-measurement rotation is specified, then three rotations are performed before measuring using a \nq{meas_rot}-command, see below.
The axes of these rotations as given in the instruction name.

The angles of the rotations are specified by the integers \nq{n\{0,1,2\}} and \nq{d\{0,1,2\}} in the same way as for single-qubit rotations.
That is, rotation \nq{i} is done by angle $\frac{\pi n_i}{2^{d_i}}$.

\textbf{Entanglement generation}

There are two commands related to entanglement generation.
A node can initiate entanglement generation with another node by using the \nq{create_epr}-command.
This command is \emph{not} blocking until entanglement has been generated but a wait-instruction (see below) can be used to block until certain a certain array has been written to, indicating that entanglement has been generated.
The remote node should also provide a \nq{recv_epr}-command.
This command does not initiate the entanglement generation but is used to provide the virtual qubit IDs that should be used for the entangled qubits.

\begin{itemize}
  \item \nq{create_epr}: Create an EPR pair with a remote node.
        \begin{enumerate}
          \item \REGISTER: Remote node ID.
          \item \REGISTER: EPR socket ID.
          \item \REGISTER: Provides the address to the array containing the virtual qubit IDs for the entangled pairs in this request.
                The value of the register should contain the address to an array with as many virtual qubit IDs stored as pair requested.
          \item \REGISTER: Provides the address to the array which holds the rest of the arguments of the entanglement generation to the network stack~\cite{dahlberg2019linklayer,kozlowski2020networklayer}.
                The value of the register should contain the address to an array with as entries as arguments in the entanglement generation request to the network stack~\cite{dahlberg2019linklayer,kozlowski2020networklayer} (except remote node ID and EPR socket ID).
          \item \REGISTER: Provides the address to the array to which information about the entanglement should be written.
                The value of the register should contain the address to an array with as many entries as $n_\mathrm{pairs} \times n_\mathrm{args}$, where $num_\mathrm{args}$ is the number of arguments in the entanglement information provided by the network stack~\cite{dahlberg2019linklayer,kozlowski2020networklayer}.
        \end{enumerate}
  \item \nq{recv_epr}: Receive an EPR pair from a remote node.
        \begin{enumerate}
          \item \REGISTER: Remote node ID.
          \item \REGISTER: EPR socket ID.
          \item \REGISTER: Provides the address to the array containing the virtual qubit IDs for the entangled pairs in this request.
                The value of the register should contain the address to an array with as many virtual qubit IDs stored as pair requested.
          \item \REGISTER: Provides the address to the array to which information about the entanglement should be written.
                The value of the register should contain the address to an array with as many entries as $n_\mathrm{pairs} \times n_\mathrm{args}$, where $n_\mathrm{args}$ is the number of arguments in the entanglement information provided by the network stack~\cite{dahlberg2019linklayer,kozlowski2020networklayer}.
        \end{enumerate}
\end{itemize}

\subsection{Waiting}\label{app:waiting}

There are three wait-commands that can wait for entries in arrays to become \emph{defined}, i.e. not \nq{null}.
Entries in a new array is by default \nq{null} (\emph{undefined}).
\begin{itemize}
  \item \nq{wait_all}: Wait for all entries in a given array slice to become not \nq{null}.
        \begin{enumerate}
          \item \textbf{ARRAY\_SLICE}: Array slice to wait for.
        \end{enumerate}
  \item \nq{wait_any}: Wait for any entry in a given array slice to become not \nq{null}.
        \begin{enumerate}
          \item \textbf{ARRAY\_SLICE}: Array slice to wait for.
        \end{enumerate}
  \item \nq{wait_single}: Wait for a single entry in an array to become not \nq{null}.
        \begin{enumerate}
          \item \textbf{ARRAY\_ENTRY}: Array entry to wait for.
        \end{enumerate}
\end{itemize}

\subsection{Deallocation}
\begin{itemize}
  \item \nq{qfree}: Stop using a qubit in the unit module.
        \begin{enumerate}
          \item \REGISTER: The virtual address of the qubit.
        \end{enumerate}
\end{itemize}

\subsection{Return}

There are two commands for returning data to the application layer.
These commands indicate that the copy of the memory on the application layer side should be updated, see above.
\begin{itemize}
  \item \nq{ret_reg}: Return a register.
        \begin{enumerate}
          \item \REGISTER: The register to return.
        \end{enumerate}
  \item \nq{ret_arr}: Return an array,
        \begin{enumerate}
          \item \ADDRESS: The address of the array to return.
        \end{enumerate}
\end{itemize}

\section{Preprocessing}
A subroutine written in text form will first be preprocessed, which does the following:
\begin{itemize}
  \item Parses preprocessing commands and handles these.
        Any preprocessing command starts with \nq{#} and should be before any command in the body of the subroutine.
        Allowed preprocessing commands are:
        \begin{itemize}
          \item \netqasm (required): Sets the \netqasm\ version in the metadata.
                \begin{nqcode}
# NETQASM 1.0\end{nqcode}
          \item \nq{APPID} (required): Sets the application ID in the metadata.
                \begin{nqcode}
# APPID 0\end{nqcode}
          \item \nq{DEFINE} (optional): Defines a macro with a key and a value.
                Any occurrence of the key prepended by \nq{\$} will be replaced with the value in the subroutine.
                Values containing spaces should be enclosed with \nq{\{\}}.

                \begin{nqcode}
# DEFINE q 0
# DEFINE add {add @0 @0 @1}\end{nqcode}
                First command replaces any occurrence of \nq{$q} with \nq{0} and second \nq{$add} with \nq{add @0 @0 @1}.
        \end{itemize}
\end{itemize}

\section{Examples}
Here we list some examples of programs written in \netqasm.
In \cref{sec:examples_netqasm}, we show some examples written directly in the \netqasm-language.
In \cref{app:examples_sdk}, we show the corresponding examples, instead written in the Python SDK.

\subsection{NetQASM}\label{sec:examples_netqasm}
\subsubsection{Classical logic (if-statement)}\label{sec:example_nq_if}
A subroutine which creates a qubit, puts in the $|+\rangle$ state, measures it and depending on the outcome performs an X-gates such that by the end of the subroutine the qubit is always in the state $|0\rangle$.
\begin{nqcode}
# NETQASM 1.0
# APPID 0
// Set the virtual qubit ID to use
set Q0 0

// Allocate and initialize a qubit
qalloc Q0
init Q0

// Perform a Hadamard gate
h Q0

// Measure the qubit
meas Q0 M0

// Branch to end if m = 0
bez M0 EXIT

// Perform X gate
x Q0

EXIT:\end{nqcode}

\subsubsection{Classical logic (for-loop)}\label{sec:example_nq_for}
A subroutine which performs a for-loop which body creates a qubit, puts in the $|+\rangle$ state and measures it. The outcomes are stored in an array.
In a higher-level language (using python syntax) the below subroutine might be written as follows:
\begin{pycode}
ms = [None] * 10

for i in range(10):
  q = Qubit()
  q.H()
  m = q.measure()
  ms[i] = m
\end{pycode}
The equivalent \netqasm subroutine is:
\begin{nqcode}
# NETQASM 1.0
# APPID 0
# DEFINE ms @0
# DEFINE i R0
# DEFINE q Q0
# DEFINE m M0
// Create an array with 10 entries (all null)
array 10 $ms

// Initialize loop counter
store $i 0

// Set the virtual qubit ID to use
set $q 0

// Loop entry
LOOP:
beq $i 10 EXIT

// Loop body
qalloc $q
init $q
h $q
meas $q $m
store $m $ms[$i]
qfree $q
add $i $i 1

// Loop exit
jmp LOOP
EXIT:\end{nqcode}

In the above subroutine \nq{DEFINE} statements have been used to clarify what registers/arrays correspond to the variables in the higher-level language example above.

\subsubsection{Create and recv EPR}\label{sec:example_nq_epr}
This code is for the side initializing the entanglement request.
\begin{nqcode}
# NETQASM 1.0
# APPID 0
# DEFINE qubits @0
# DEFINE args $1
# DEFINE entinfo @2
// Initilizer side

// Setup array with virtual qubit IDs to be used
// for the EPR pairs
array 1 $qubits
store 0 $qubits[0]

// Setup array to store other arguments to entanglement
// generation request
array 20 $args

// Setup array to store entanglement information
array 10 $entinfo

// Create entanglement
// Remote node ID 0 and EPR socket ID 0
// NOTE that these IMMEDIATEs will be replaced by
// REGISTERs when pre-processing.
create_epr 1 0 $qubits $args $entinfo

// Wait for the entanglement to succeed
// i.e. that all entries in the entinfo array becomes
// valid.
wait_all $entinfo[0:10]

// Measure the entanglement qubit
load Q0 $qubits[0]
meas Q0 M0

// Return the outcome
ret_req M0\end{nqcode}

This code is for the receiving side.
\begin{nqcode}
# NETQASM 1.0
# APPID 0
# DEFINE qubits @0
# DEFINE entinfo @1
// Receiver side (very similar to the initializer side)

// Setup array with virtual qubit IDs to be used
// for the EPR pairs
array 1 $qubits
store 0 $qubits[0]

# Setup array to store entanglement information
array 10 $entinfo

// Receive entanglement
// Remote node ID 1 and EPR socket ID 0
// NOTE that these IMMEDIATEs will be replaced by
// REGISTERs when pre-processing.
recv_epr 1 0 $qubits $entinfo

// Wait for the entanglement to succeed
wait_all $entinfo[0:10]

// Measure the entanglement qubit
load Q0 $qubits[0]
meas Q0 M0

// Return the outcome
ret_req M0\end{nqcode}

\subsection{SDK}\label{app:examples_sdk}
Each of the examples in this section are functionally the same as the examples in section~\ref{sec:examples_netqasm}.
A compiler will produce a similar subroutine as the examples in the previous section but might vary depending on the exact implementation of the compiler.

\subsubsection{Classical logic (if-statement)}
Functionally the same as the \netqasm-subroutine (\cref{sec:example_nq_if}).
\begin{pycode}
# Setup connection to backend
# as the node Alice
with NetQASMConnection("Alice") as alice:
  # Create a qubit
  q = Qubit(alice)
  # Perform a Hadamard on the qubit
  q.H()
  # Measure the qubit
  m = q.measure()
  # Conditionally apply a X-gate
  with m.if_eq(1):
    q.X()
\end{pycode}

\subsubsection{Classical logic (for-loop)}
Functionally the same as the \netqasm-subroutine (\cref{sec:example_nq_for}).
\begin{pycode}
# Setup connection to backend
# as the node Alice
with NetQASMConnection("Alice") as alice:
  # Create an array for the outcomes
  outcomes = alice.new_array(10)
  # For-loop
  with alice.loop(10) as i:
    # Create a qubit
    q = Qubit(alice)
    # Perform a Hadamard on the qubit
    q.H()
    # Measure the qubit
    m = q.measure()
    # Add the outcome to the array
    outcomes[i] = m
\end{pycode}

\subsubsection{Create and recv EPR}
Functionally the same as the \netqasm-subroutine (\cref{sec:example_nq_epr}).

This code is for the side initializing the entanglement request.
\begin{pycode}
# Setup an EPR socket with the node Bob
epr_socket = EPRSocket("Bob")
# Setup connection to backend
# as the node Alice
with NetSquidConnection(
"Alice",
epr_sockets=[epr_socket],
):
  # Create entanglement
  epr = epr_socket.create()[0]
  # Measure the entangled qubit
  m = epr.measure()
\end{pycode}

This code is for the receiving side.
\begin{pycode}
# Setup an EPR socket with the node Alice
epr_socket = EPRSocket("Bob")
# Setup connection to backend
# as the node Bob
with NetSquidConnection(
"Alice",
epr_sockets=[epr_socket]
):
  # Create entanglement
  epr = epr_socket.recv()[0]
  # Measure the entangled qubit
  m = epr.measure()
\end{pycode}

\section{Simulation details}\label{app:simulation}

In this section we detail how simulations in \cref{sec:evaluation} were
performed and what models and parameters were used. All simulations used the
\netqasm\ SDK~\cite{git_netqasm}, using
\netsquid~\cite{netsquid,coopmans2021netsquid} as the underlying simulator. All
code used in these simulations can also be found at~\cite{git_squidasm}.

\subsection{Noise model}
In both the teleportation and the blind quantum computing scenario we used the
same model for nitrogen-vacancy centres in diamonds as was used
in~\cite{dahlberg2019linklayer} and~\cite{coopmans2021netsquid}. All gates
specified by the application in the SDK were translated to NV-specific gates,
see \cref{tab:gates}, using a simple compiler without any optimization. The
parameters used in the model from~\cite{dahlberg2019linklayer} are listed in
\cref{tab:gates,tab:noise}, together with an explanation and a reference.
\nq{ec_controlled_dir_xy} are the native two-qubit gates of the NV-platform,
ideally performing one of the unitary operations
\begin{align}
  U_\mathrm{ec_x}(\alpha) & = \begin{pmatrix}R_x(\alpha) & 0 \\ 0 & R_x(-\alpha) \end{pmatrix} \label{eq:crot_x} \\
  U_\mathrm{ec_y}(\alpha) & = \begin{pmatrix}R_y(\alpha) & 0 \\ 0 & R_y(-\alpha) \end{pmatrix} \label{eq:crot_y} \\
\end{align}
where $R_x(\alpha)$ and $R_y(\alpha)$ are the rotation matrices around $X$ and
$Y$, respectively. When sweeping the duration and noise of this two-qubit gate
the same value is also used for the \nq{carbon_xy_rot} ($X$- and $Y$-rotations
on the carbon) on the storage qubits, since these are also effectively done with
a similar operation also involving the communication qubit (electron). All noise
indicated by a fidelity in \cref{tab:noise} are applied as depolarising noise by
applying the perfect operation, producing the state $\rho_\mathrm{ideal}$, and
mapping this to
\begin{equation}\label{eq:depolarising}
  \rho_\mathrm{noisy}=(1-p)\rho_\mathrm{ideal} + \frac{p}{3}X\rho_\mathrm{ideal}X + \frac{p}{3}Y\rho_\mathrm{ideal}Y + \frac{p}{3}Z\rho_\mathrm{ideal}Z
\end{equation}
where $X$, $Y$ and $Z$ are the Pauli operators in
\cref{eq:pauli_x,eq:pauli_y,eq:pauli_z}, $p=\frac{4}{3}(1 - F)$, with $F$ being
the value specific in \cref{tab:noise}. Decoherence noise is specific as $T_1$
(energy/thermal relaxation time) and $T_2$ (dephasing time)~\cite{Nielsen2010}.

\begin{table}
  \centering
  \begin{tabular}{||c|c|c|c||}
    \hline
    Gate                      & Durations (ns) & Explanation                                                \\
    \hline\hline
    \nq{electron_init}        & 2e3            & Initialize a communication qubit (electron) to $|0\rangle$ \\
    \nq{electron_rot}         & 5              & single-qubit rotation on communication qubit (electron)    \\
    \nq{measure}              & 3.7e3          & Measure communication qubit (electron)                     \\
    \nq{carbon_init}          & 3.1e5          & Initialize a storage qubit (carbon) to $|0\rangle$         \\
    \nq{carbon_xy_rot}        & $t$            & $X$/$Y$-rotation on storage qubit (carbon)                 \\
    \nq{carbon_z_rot}         & 5              & $Z$-rotation on storage qubit (carbon)                     \\
    \nq{ec_controlled_dir_xy} & $t$            & Native two-qubit gates, see \cref{eq:crot_x,eq:crot_y}     \\
    \hline
  \end{tabular}
  \caption{
    Gate durations for scenario \textbf{B} of \cref{sec:evaluation}.
    $t$ is the value being swept in \cref{fig:sweep_gate_time}.
    All values are from~\cite{dahlberg2019linklayer}.
  }\label{tab:gates}
\end{table}

\begin{table}
  \centering
  \begin{tabular}{||c|c|c|c||}
    \hline
    Parameter                 & Value        & Explanation                                                 \\
    \hline\hline
    \nq{electron_T1}          & 1 hour       & $T_1$ of communication qubit (electron)                     \\
    \nq{electron_T2}          & 1.46 seconds & $T_2$ of communication qubit (electron)                     \\
    \nq{electron_init}        & 0.99)        & Fidelity to initialize communication qubit (electron)       \\
    \nq{electron_rot}         & 1.0          & Fidelity for $Z$-rotation on communication qubit (electron) \\
    \nq{carbon_T1}            & 10 hours     & $T_1$ of storage qubit (carbon)                             \\
    \nq{carbon_T2}            & 1 second     & $T_2$ of storage qubit (carbon)                             \\
    \nq{carbon_init}          & 0.997        & Fidelity to initialize storage qubit (carbon)               \\
    \nq{carbon_z_rot}         & 0.999        & Fidelity for $Z$-rotation on storage qubit (carbon)         \\
    \nq{carbon_xy_rot}        & $f$          & Fidelity for $X$/$Y$-rotation on storage qubit (carbon)     \\
    \nq{ec_controlled_dir_xy} & $f$          & Fidelity for native two-qubit gate                          \\
    \nq{prob_error_meas_0}    & 0.05         & Probability of flipped measurement outcome for $|0\rangle$  \\
    \nq{prob_error_meas_1}    & 0.005        & Probability of flipped measurement outcome for $|1\rangle$  \\
    \nq{link_fidelity}        & 0.9          & Fidelity of generated entangled pair.                       \\
    \hline
  \end{tabular}
  \caption{
    Noise parameters for used in the simulations of \cref{sec:evaluation}.
    $f$ is the value being swept in \cref{fig:sweep_gate_noise} and \cref{fig:plot_bqc}.
    All fidelities are realized by a applying depolarising noise as in \cref{eq:depolarising}.
    All values are from~\cite{coopmans2021netsquid}, except \nq{link_fidelity} which is set to relatively high value to avoid this being the major noise-contribution and preventing any conclusions to be made.
  }\label{tab:noise}
\end{table}

\subsection{BQC application and flavors}
In \cref{sec:evaluation-flavours} we simulated the blind quantum computation
(BQC) application from \cref{fig:bqc_app}. The code for this is available at
\cite{git_squidasm}.

In the scenario when the application code was compiled to subroutines with the
vanilla lavour, the \QNPU had to map the vanilla instructions to NV-native
operations on the fly. We used the gate mappings listed below. For
convenience we use \nq{PI} and \nq{PI_OVER_2} for $\pi$ and $\frac{\pi}{2}$
respectively.

A \nq{h} (Hadamard) vanilla instruction was mapped to the following NV instruction sequence:
\begin{nqcode}
  rot_y PI_OVER_2
  rot_x PI\end{nqcode}

A \nq{cnot C S} vanilla instruction between a communication qubit (C) and a storage
qubit (S) (as specified in the unit module) was mapped to the following NV
instruction sequence:
\begin{nqcode}
  cx_dir C S PI_OVER_2
  rot_z C -PI_OVER_2
  rot_x S -PI_OVER_2\end{nqcode}

A \nq{cnot S C} vanilla instruction between a store qubit (S) and a communication
qubit (C) (as specified in the unit module) was mapped to the following NV
instruction sequence:
\begin{nqcode}
  rot_y C PI_OVER_2
  rot_x C PI
  rot_y S PI_OVER_2
  cx_dir C S PI_OVER_2
  rot_z C -PI_OVER_2
  rot_x S -PI_OVER_2
  rot_y S PI_OVER_2
  rot_y C PI_OVER_2
  rot_x C PI\end{nqcode}

A \nq{cphase C S} vanilla instruction between a communication qubit (C) and a storage
qubit (S) (as specified in the unit module) was mapped to the following NV
instruction sequence:
\begin{nqcode}
  rot_y S PI_OVER_2
  cx_dir C S PI_OVER_2
  rot_z C -PI_OVER_2
  rot_x S -PI_OVER_2
  rot_y S -PI_OVER_2\end{nqcode}

\end{document}